\documentclass[journal]{IEEEtran}

\usepackage{cite}
\usepackage{algorithm}

\ifCLASSINFOpdf
   \usepackage[pdftex]{graphicx}
\else
   \usepackage[dvips]{graphicx}
   \DeclareGraphicsExtensions{.eps}
\fi
\usepackage{algorithmic}
\usepackage{amssymb}
\newcounter{MYtempeqncnt}

\usepackage{dblfloatfix}
\usepackage{url}

\usepackage{color}
\usepackage{amsmath,epsfig,subfigure,paralist,verbatim}
\usepackage{amsfonts}
\usepackage{epsfig,psfrag}
\usepackage{dsfont}
\usepackage{epstopdf}

\newcommand{\norm}[1]{\lVert#1\rVert}
\newcommand{\tindx}{t}
\newcommand{\Tindxter}{T}
\newcommand{\nindx}{n}

\newcommand{\probe}{p}
\newcommand{\response}{x}

\newcommand{\utility}{u}
\newcommand{\budget}{I}
\newcommand{\dataset}{\mathcal{D}}
\newcommand{\setresponse}{X}
\newcommand{\potfun}{V}
\newcommand{\estpotfun}{\hat{V}}

\newcommand{\noise}{w}
\newcommand{\obsdataset}{\mathcal{D}_\text{obs}}
\newcommand{\obsresponse}{y}

\newcommand{\vecx}{\ensuremath{x}}

\newcommand{\reals} {\Bbb{R}}
\newcommand{\p}{\prime}
\newcommand{\qed} {{$\hfill\blacksquare$}}

\newtheorem{definition}{Definition}[section]

\newtheorem{remark}{Remark}[section]
\newtheorem{theorem}{Theorem}[section]

\newtheorem{lemma}{Lemma}[section]
\newtheorem{corollary}{Corollary}[section]

\begin{document}
%
\title{Reinforcement Learning and Nonparametric Detection of Game-Theoretic Equilibrium Play in Social Networks}
%
%
%

\author{Omid Namvar Gharehshiran,
		William Hoiles,~\IEEEmembership{Student Member,~IEEE},
		Vikram~Krishnamurthy,~\IEEEmembership{Fellow,~IEEE}
\thanks{O. N. Gharehshiran, W. Hoiles, and V. Krishnamurthy are with the department of Electrical and Computer
Engineering, University of British Columbia, Vancouver, V6T 1Z4, Canada (e-mail: omidn@ece.ubc.ca; whoiles@ece.ubc.ca; vikramk@ece.ubc.ca).
This research was supported by the NSERC Strategic grant and the Canada Research Chairs program.}
}




\maketitle

\begin{abstract}
This paper studies two important signal processing aspects of equilibrium behavior in non-cooperative games arising in social networks, namely, \emph{reinforcement learning} and \emph{detection} of equilibrium play. The first part of the paper presents a reinforcement learning (adaptive filtering) algorithm that facilitates learning an equilibrium by resorting to diffusion cooperation strategies in a social network. Agents form homophilic social groups, within which they exchange past experiences over an undirected graph. It is shown that, if all agents follow the proposed algorithm, their global behavior is attracted to the correlated equilibria set of the game. The second part of the paper provides a test to detect if the actions of agents are consistent with play from the equilibrium of a concave potential game. The theory of revealed preference from microeconomics is used to construct a non-parametric decision test and statistical test which only require the probe and associated actions of agents.  A stochastic gradient algorithm is given to optimize the probe in real time to minimize the Type-II error probabilities of the detection test subject to specified Type-I error probability. We provide a real-world example using the energy market, and a numerical example to detect malicious agents in an online social network. 
\end{abstract}


%


\newcommand{\plyrset}{\mathcal{K}}
\newcommand{\actset}{\mathcal{A}}
\newcommand{\inertia}{\mu}
\newcommand{\plyrind}{k}
\newcommand{\fact}{A}
\newcommand{\utilityk}{u^\plyrind}
\newcommand{\mixedstrat}{\mathbf{p}}
\newcommand{\regmatplyr}{R^\plyrind}
\newcommand{\act}{a}
\newcommand{\dtimee}{n}
\newcommand{\dtimenext}{\dtimee+1}
\newcommand{\mixedstratind}{p}
\newcommand{\regmatplyrdiff}{\bar{R}^\plyrind}
\newcommand{\stepsize}{\varepsilon}
\newcommand{\actprof}{\mathbf{a}}
\newcommand{\neighborhoodk}{\mathcal{N}^\plyrind}
\newcommand{\cneighborhoodk}{\mathcal{N}^\plyrind_c}
\newcommand{\weightkl}{w_{\plyrind l}}
\newcommand{\indicatori}{I\left( \act^\plyrind_\dtimee = i\right)}
\newcommand{\indicatorj}{I\left( \act^\plyrind_\dtimee = j\right)}
\newcommand{\fplyr}{K}
\newcommand{\actk}{a^\plyrind}
\newcommand{\communitys}{C^s}
\newcommand{\fcom}{S}
\newcommand{\comind}{s}
\newcommand{\comset}{\mathcal{S}}
\newcommand{\globbehav}{\mathbf{z}}
\newcommand{\unitvec}{\mathbf{e}}
\newcommand{\lyap}{V}
\newcommand{\gamevertex}{V}
\newcommand{\gameedge}{\mathcal{E}}
\newcommand{\regplyrij}{r^\plyrind_n(i,j)}
\newcommand{\ctimeet}{(t)}
\newcommand{\regmatplyrinterpol}{R^{\plyrind,\stepsize}\ctimeet}
\newcommand{\regmatplyrinterpolcdot}{R^{\plyrind,\stepsize}(\cdot)}
\newcommand{\markovgame}{\theta}
\newcommand{\statespacegame}{\mathcal{M}}
\newcommand{\fstate}{M}
\newcommand{\markovspeed}{\rho}
\newcommand{\contmarkovmat}{Q}
\newcommand{\contmarkovmatind}{q_{ij}}
\newcommand{\globalregret}{\mathbf{R}}
\newcommand{\globalregretinterpol}{\mathbf{R}^{\stepsize}}
\newcommand{\statdistgroupnor}{\boldsymbol{\sigma}^\plyrind}
\newcommand{\statdistindgroupind}{\sigma_i^\plyrind\big(\regmatplyr\big)}
\newcommand{\statdistindgroupnor}{\sigma_i^\plyrind}
\newcommand{\statdistk}{\boldsymbol{\psi}^{\plyrind}}
\newcommand{\statdistind}{\psi}
\newcommand{\statdistindi}{\psi^{\plyrind}_i}
\newcommand{\statdistindj}{\psi^{\plyrind}_j}
\newcommand{\diffinclglobal}{\mathbf{H}}
\def\mset{\mathcal{M}}
\def\lb{\left[}
\def\rb{\right]}
\def\lbr{\left\lbrace}
\def\rbr{\right\rbrace}
\def\bar{\overline}
\def\cd{(\cdot)}
\def\M{\mathcal{M}}
\def\G{\mathcal{G}}
\def\RR{\mathbb{R}}
\def\C{\mathcal{C}}
\def\B{\mathcal{B}}
\def\explor{\delta}
\def\CEdistance{\epsilon}
\def\ee{\mathbb{E}}
\def\z{\mathbf{z}}
\def\e{\mathbf{e}}
\def\x{\mathbf{x}}

\section{Introduction}
\label{sec:intro}
\IEEEPARstart{L}{earning}, rationalizability, and equilibrium in games are of central importance in the analysis of social networks.
Game theory has traditionally been used in economics and social sciences with a focus on fully rational interactions where strong assumptions are made on the information patterns available to individual agents. In comparison, social networks are comprised of agents with limited cognition and communication capabilities, and it is the dynamic interactions among agents that are of interest. This, together with the interdependence of agents' choices, motivates the need for game-theoretic learning models for agents interacting in social network. 

The game-theoretic notion of equilibrium describes a condition of global coordination where all agents are content with their social welfare. Reaching an equilibrium, however, involves a complex process of agents guessing what each other will do. 
Game-theoretic learning explains how such coordination might arise as a consequence of a long-run process of learning from interactions and adapting behavior~\cite{FL08}. 


\subsection{Main Ideas and Organization}

\begin{figure}[!t]
	\setlength{\abovecaptionskip}{0em}
	\setlength{\belowcaptionskip}{0em}
	\vspace{-0.3cm}
	\begin{center}
		\includegraphics[width=0.45\textwidth]{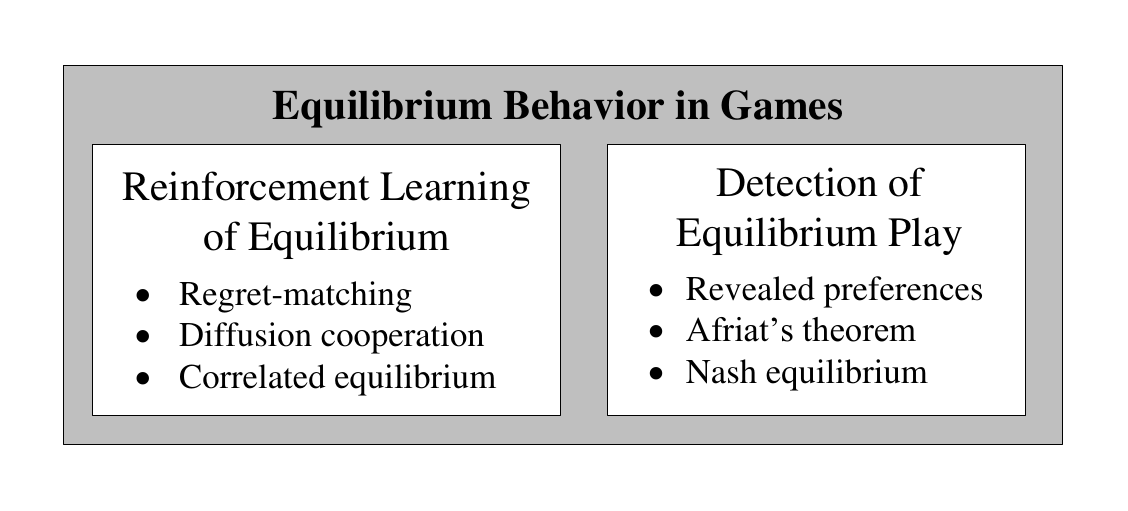}
	\end{center}
	\vspace{-0.3cm}
	\caption{Two aspects of game-theoretic equilibrium behavior in social networks discussed in this paper. Both aspects involve multi-agent signal processing over a network. Stochastic approximation algorithms are used in both cases to devise the desired scheme.
	}
	\label{fig:intro}
\end{figure}

The two aspects of equilibrium behavior in games that are addressed in this paper along with the main tools used are illustrated in Fig.~\ref{fig:intro}.  These two aspects are relevant to the broad area of machine learning of equilibria in social networks. The main results of this paper are summarized below:

\subsubsection{Reinforcement learning dynamics in social networks}
The first part of this paper (Sec.~\ref{sec:game} to Sec.~\ref{sec:main-results-game}) addresses the questions: Can a social network of self-interested agents that possess limited sensing and communication capabilities reach a global equilibrium behavior in a distributed fashion? 
If so, can formation of social groups that exhibit identical homophilic characteristics facilitate the learning dynamics within the network? The main idea 
is to propose a diffusion based stochastic approximation algorithm (learning scheme) that if each agent deploys,
the collective behavior of the social network converges to a correlated equilibrium.

Sec.~\ref{subsec:game-formulation}
introduces non-cooperative games with homophilic social groups in a social network. Homophily\footnote{In~\cite{ST11} the following illustrative example is provided for homophily behavior in social networks: ``If your friend jumped off a bridge, would you jump too?'' A possible reasons for answering ``yes'' is that you are friends as a result of your fondness for jumping off bridges. Notice that this is different to contagion behavior where ``your friend inspired you to jump off the bridge''. Due to space restrictions we do not consider contagion behavior in this paper.\label{footnote:1}} refers to a tendency of various types of individuals to exchange information with others who are similar to themselves. The detection of social groups that show common behavioural characteristics can be performed using such methods as matched sample estimation~\cite{AMS09}. 
Sec.~\ref{subsec:CE} introduces correlated equilibrium~\cite{Aum87} as the solution concept for such games. 
Correlated equilibrium is a generalization of Nash equilibrium, however, is more realistic in multi-agent learning scenarios since observation of the past history of decisions (or their associated outcomes) naturally correlates agents' future decisions. 

%
%
%

In Sec.~\ref{sec:algorithm} we present a regret-based reinforcement learning algorithm that, resorting to diffusion cooperation strategies~\cite{Say14,Say14b}, implements cooperation among members of a social group. The proposed algorithm is based on the well-known regret-matching algorithm~\cite{HM00,HM01a,HMB13}.
The proposed algorithm\footnote{Although we use the term ``algorithm,'' the learning procedure mimics human behavior; it involves minimizing a moving average regret and random experimentation.} suits the emerging information patterns in social networks, and allows to combine the past experiences of agents across the network, which facilitates the learning dynamics and enables agents to respond in real time to changes underlying the network. To the best of our knowledge, this is the first work that uses diffusion cooperation strategies to implement collaboration in game-theoretic reinforcement learning. Sec.~\ref{sec:main-results-game} shows that if each agent individually follows the proposed algorithm, the  experienced regret will at most be $\epsilon$ after sufficient repeated plays of the game. Moreover, if all agents follow the proposed algorithm independently, their collective behavior across the network will converge to an $\epsilon$-distance of the polytope of correlated equilibria. 


\subsubsection{Detection of equilibrium play in a social networks}
The second part of the paper (Sec.~\ref{sec:revealed} to Sec.~\ref{sec:ExamplesofEquilibriumPlay}) addresses the question: Given datasets of the external influence and actions of agents in a social network, is it possible to detect if the behavior of agents is consistent with play from the equilibrium of a concave potential game. The theory of revealed preference from microeconomics is used to construct a non-parametric decision test which only requires the time-series of data $\dataset=\{(\probe_\tindx,\response_\tindx): \tindx\in\{1,2,\dots,\Tindxter\}\}$ where $\probe_\tindx\in\mathds{R}^m$ denotes the external influence, and $\response_\tindx\in\mathds{R}^m$ denotes the action of an agent. These questions are fundamentally different to the {\em model-based} theme that is widely used in the signal processing literature in which an objective function (typically convex) is proposed and then algorithms are constructed to compute the minimum. In contrast, the revealed preference approach is {\em data-centric}---we wish to determine whether the dataset is obtained from the interaction of utility maximizers.

Sec.~\ref{sec:revealed} introduces how revealed preferences can be used to detect if the actions of agents originated from play from a concave potential game using only the external influence and actions of the agents. Specifically in Sec.~\ref{subsec:afriatstheorem} we introduce the preliminary tools of revealed preference for detecting utility maximization of single agents. Sec.~\ref{subsec:detectiontests} provides a non-parametric test to detect if the actions of agents is a result of play from a concave potential game. 
If the actions are measured in noise then Sec.~\ref{subsec:FeasibilityTestforNashRationality} provides a non-parametric statistical test for play from a concave potential game with guaranteed Type-I error probability. To reduce the probability of Type-II errors of the statistical test Sec.~\ref{subsec:SPSA} provides a simultaneous perturbation stochastic gradient (SPSA) algorithm to adjust the external influence in real-time. Two examples are provided to illustrate the application decision test, statistical test, and SPSA algorithm developed in Sec.~\ref{subsec:detectiontests} to Sec.~\ref{subsec:SPSA}.

Concave potential games are considered for the detection test as the detection tests for $\dataset$ to satisfy a Nash equilibrium are very weak~\cite{Deb09}. In this paper the requirement of $\dataset$ to be consistent with Nash equilibrium of a concave potential game provides stronger restrictions when compared to only a Nash equilibrium while still encompassing a large set of utility functions. 

An interesting aspect of both parts of this paper is the ordinal nature of decision making. In the learning dynamics of Sec.~\ref{sec:algorithm}, the actions taken are ordinal; in the data-set parsing of Sec.~\ref{sec:revealed}, the utility function obtained is ordinal. Humans make ordinal decisions\footnote{\label{footnote:ordinal}Humans typically convert numerical attributes to ordinal scales before making  decisions. For example,
it does not matter if the cost of a meal at a restaurant is \$200 or \$205; an individual would classify this cost as ``high". Also credit rating agencies use ordinal symbols such as AAA, AA, A.} since humans tend to think in symbolic ordinal terms.

\subsection{Literature}

Game theoretic models for social networks have been studied widely~\cite{JW96,GS12}.
For example,~\cite{K07} formulate graphical games model where each agent's influence is restricted to its immediate neighbors.
The reader is referred to~\cite{KNH14} for an treatment of interactive sensing and decision making in social networks from the signal processing perspective.

\subsubsection*{Reinforcement Learning Dynamics}
%
Regret-matching~\cite{HM00,HM01a,HMB13}
is known to guarantee convergence to the set of correlated equilibria~\cite{Aum87} under no structural assumptions on the game model. The correlated equilibrium arguably provides a natural way to capture conformity to social norms~\cite{CW12}. It can be interpreted as a mediator~\cite{XV12} instructing people to take actions according to some commonly known probability distribution. 
The regret-based adaptive procedure in~\cite{HM00} assumes a fully connected network topology, whereas the regret-based reinforcement learning algorithm in~\cite{HM01a} assumes a set of isolated agents who neither know the game, nor share information of their past decisions with others.
In~\cite{NKY13a}, a regret-matching algorithm is developed when agents exchange information over a non-degenerate network connectivity graph. 


The cooperation strategies in adaptive networked systems can be classified as: (i) \emph{incremental strategies}~\cite{Ber97}, (ii) \emph{consensus strategies}~\cite{BT97},
and (iii) \emph{diffusion strategies}~\cite{Say14}. In the first class of strategies, information is passed from one agent to the next over a cyclic path until all agents are visited. In contrast, in the latter two, cooperation is enforced among multiple agents rather than between two adjacent neighbors. Diffusion strategies are shown to outperform consensus strategies in~\cite{tu2012diffusion}; therefore, we concentrate on former to implement cooperation among social agents in Sec.~\ref{sec:algorithm}. The diffusion strategies have been used previously for distributed estimation, distributed decision making, and distributed Pareto optimization in adaptive networks~\cite{Say14b}.

\subsubsection*{Detection of Equilibrium Play}
Humans can be viewed as social sensors that interact over a social network  to  provide information about their environment. Social sensors go beyond physical sensors--for example, user preferences for a particular movie are available on Rotten Tomatoes but are difficult to measure via physical sensors. Social sensors  present unique challenges from a statistical estimation point of view. First, social sensors  interact with and influence other social sensors. Second, due to privacy concerns and time-constraints, social sensors typically do not reveal their personal preference rankings between actions. In classical revealed preference theory in the micro-economics literature Afriat's theorem gives a non-parametric finite sample test to decide if an agent's actions to an external influence are consistent with utility maximization~\cite{Afr67}. The revealed preference test for single agents has been applied to measuring the welfare effect of price discrimination, analyzing the relationship
between prices of broadband Internet access and time of use service, and auctions for advertisement position placement on page search results from Google \cite{Var12}.

For interacting agents in a social network (i.e. players in a game), single agent tests are not suitable. Typically the study of interacting agents in a game require parametric assumptions on the form of the utility function of the agents. Deb~\cite{Deb09} was the first to propose a detection test for players engaged in a concave potential game based on Varian's and Afriat's work~\cite{Var83,Afr67}. Potential games were introduced by Monderer and Shapley~\cite{MS96} and are used extensively in the literature to study the strategic behaviour of utility maximization agents. A classical example is the {\it congestion game}~\cite{RR73} in which the utility of each agent depends on the amount of resource it and other agents in the social network use. Recently the analysis of energy use scheduling and demand side management schemes in the energy market was performed using potential games~\cite{CVH13}.

\begin{figure*}[!hb]
	\hrulefill
	\vspace*{-2pt}
	\setcounter{MYtempeqncnt}{\value{equation}}
	\setcounter{equation}{3}
	\begin{equation}
	\label{eq:CE_defn}
	\textstyle\mathcal{Q}_\epsilon = \Big\{ \boldsymbol{\pi}:
	\sum_{\actprof^{-k}}\pi^k\big(i, \actprof^{-k}\big)\left[u^{k}\big(j, \actprof^{-k}\big) - u^k\big(i, \actprof^{-k}\big)\right] \leq \epsilon, \;\forall i,j\in\actset^k, k\in\plyrset \Big\}
	\end{equation}
	\setcounter{equation}{\value{MYtempeqncnt}}
	\vspace*{-30pt}
\end{figure*}

\section{Learning Equilibria in Non-Cooperative Games With Homophilic Social Groups}
\label{sec:game}
This section introduces a class of non-cooperative games with homophilic social groups.
Homophily refers to a tendency of various types of individuals to associate with others who are similar to themselves---see footnote~\ref{footnote:1}. Agents in homophilic relationships share common characteristics that motivates their communication. 
We then proceed to present and elaborate on a prominent solution concept in non-cooperative games, namely, correlated equilibrium.

\subsection{Non-Cooperative Game Model}
\label{subsec:game-formulation}
The standard representation of a non-cooperative game, 
known as \emph{normal form} or \emph{strategic form} game
is comprised of three elements:

\emph{1. Set of agents:} $\plyrset = \lbr 1,\cdots,\fplyr\rbr$. Essentially, an agent models an entity that is entitled to making decisions.
Agents may be people, sensors, mobile devices, etc., and are indexed by $\plyrind\in\plyrset$.

\emph{2. Set of actions:} $\actset^\plyrind = \lbr 1,\cdots,A^k\rbr$, that denotes the actions, also referred to as \emph{pure strategies}, available to agent $\plyrind$ at each decision point. A generic action taken by agent $\plyrind$ is denoted by $\actk$. The actions of agents may range from deciding to establish or abolish links with other agents~\cite{BG00} to choosing among different technologies~\cite{ZW03}.

A generic joint \emph{action profile} of all agents is denoted by
\begin{displaymath}
\actprof = \big(\act^1,\cdots,\act^\fplyr\big)\in \actset^\plyrset, \;\;\mbox{where}\;\;\actset^\plyrset = \actset^1\times\cdots\times\actset^\fplyr,
\end{displaymath}
and $\times$ denotes the Cartesian product. Following the common notation in game theory, one can rearrange $\actprof$ as
\begin{displaymath}
\actprof = \big(\actk,\actprof^{-\plyrind}\big), \;\mbox{where}\;\;
\actprof^{-\plyrind} = \big( \act^1,\cdots, \act^{\plyrind-1}, \act^{\plyrind+1}, \cdots,\act^\fplyr\big)
\end{displaymath}
denotes the action profile of all agents excluding agent $\plyrind$.

\emph{3. Utility function:} $\utilityk:\actset^\plyrset\to\RR$ is bounded, and determines the payoff to agent $\plyrind$ as a function of the action profile $\actprof$ taken by all agents. The interpretation of such a payoff is the aggregated rewards and costs associated with the chosen action as the outcome of the interaction. The payoff function can be quite general: It could reflect reputation or privacy, using the models in~\cite{GGG09,Mui02}, or benefits and costs associated with maintaining links in a social network, using the models in~\cite{BG00,ZV13b}. It could also reflect benefits of consumption and the costs of production, download, and upload in content production and sharing over peer-to-peer networks~\cite{GLML01}, or the capacity available to users in communication networks~\cite{JLL12}.

Throughout the paper, we restrict our attention to non-cooperative games in social networks in which agents have identical homophilic characteristics. These situations are modeled by a \emph{symmetric non-cooperative game} in the economics literature, that is formally defined as follows:
\vspace{0.1cm}
\begin{definition}
	\label{def:symmetric-game}
	A normal-form non-cooperative game is \emph{symmetric} if agents have identical action spaces, i.e., $\actset^\plyrind = \actset = \{1,\ldots,A\}$ for all $k\in\plyrset$, and for all $\plyrind, l\in\plyrset$:
	\begin{equation}
	\label{eq:symmetric-game}
	\utilityk\left(\actk,\actprof^{-\plyrind}\right) = u^{l}\left(\act^l,\actprof^{-l}\right),\;\mbox{if}\; 
	\actk = \act^l,\;\;\actprof^{-\plyrind} = \actprof^{-l}.
	\end{equation}
\end{definition}
\vspace{0.1cm}
Intuitively speaking, in a symmetric game, 
the identities of agents can be changed without transforming the payoffs associated with decisions. Symmetric games have been used in the literature to model interaction of buyers and sellers in the global electronic market~\cite{BWZ00}, clustering~\cite{Pel09}, cooperative spectrum sensing~\cite{WLC08}, and network formation models with costs for establishing links~\cite{MN00}.
%

\begin{figure*}[!hb]
	\vspace*{2pt}
	\setcounter{MYtempeqncnt}{\value{equation}}
	\setcounter{equation}{4}
	\begin{align}
	\label{eq:regret}
	r^\plyrind_{\dtimenext}(i,j) = \regplyrij + \stepsize\lb \frac{\mixedstratind^\plyrind_\dtimee(i)}{\mixedstratind^\plyrind_\dtimee(j)}
	\utilityk_\dtimee\left(\actk_\dtimee\right)\cdot \indicatorj - \utilityk_\dtimee\left(\actk_\dtimee\right)\cdot\indicatori - \regplyrij\rb
	\end{align}
	\setcounter{equation}{\value{MYtempeqncnt}}
	\vspace*{-30pt}
\end{figure*}

Agents can further subscribe to social groups within which they share information about their past experiences. This is referred to as \emph{neighborhood monitoring}~\cite{NKY13a}.
The communication among agents, hence, their level of ``social knowledge,'' can be captured by a connectivity graph, defined as follows:
\vspace{0.1cm}
\begin{definition}
	\label{def:connectivity-graph}
	The \emph{connectivity graph} is a simple\footnote{A simple graph is an unweighted, undirected graph containing no self loops or multiple edges.} graph $\G = (\gameedge,\plyrset)$, where agents form vertices of the graph, and
	\begin{equation}
	(\plyrind, l) \in \gameedge \Leftrightarrow 
	\begin{array}{c}
	\mbox{Agents $k$ and $l$ exchange information.}
	\end{array}\nonumber
	\end{equation}
	The \emph{open} and \emph{closed neighborhoods} of each agent $\plyrind$ are then, respectively, defined by
	\begin{equation}
	\label{eq:neighbors}
	\neighborhoodk := \lbr l\in\plyrset ; (\plyrind,l)\in\gameedge\rbr,\;\;\mbox{and}\;\;
	\cneighborhoodk := \neighborhoodk \cup \lbr \plyrind\rbr.
	\end{equation}
\end{definition}
\vspace{0.1cm}

Agents are, in fact, oblivious to the existence of other agents except their immediate neighbors on the network topology, nor are they aware of the dependence of the outcome of their decisions on those of other agents outside their social group. Besides exchanging past decisions with neighbors, agents realize the stream of payoffs as the outcome of their choices. At each time $n=1,2,\ldots$, each agent $k$ makes a decision $\actk_n$, and realizes her utility
\begin{equation}
\label{eq:realized-payoff}
\utilityk_n\left(\actk_n\right) = \utilityk\left(\actk_n,\actprof^{-k}_n\right).
\end{equation}
Here, we assume that the agents are unaware of the exact form of the utility functions. However, even if agent $k$ knows the utility function, computing utilities is impossible as she observes some (but not all) elements of $\actprof^{-k}_n$.

\begin{remark}
	It is straightforward to generalize the game model described above to social networks in which agents form multiple homophilic groups within which each agent forms a social group of its own.
	The algorithm that we present next can be employed in such clustered networks with no further modification. However, for simplicity of the presentation, we continue to use the single homophilic group in the rest of this paper.
\end{remark}

\subsection{Correlated Equilibrium}
\label{subsec:CE}
In the first part, we focus on correlated equilibrium, which is defined as follows:
\vspace{0.1cm}
\begin{definition}[Correlated Equilibrium]
	\label{def:correlated_eq}
	Let $\boldsymbol{\pi}$ denote a joint distribution on the joint action space $\actset^\plyrset$, i.e., $$\textstyle\pi\left(\actprof\right)\geq 0,\; \forall \actprof\in\actset^\plyrset,\; \mbox{and}\;\sum_{\actprof\in\actset^\plyrset}\pi\left(\actprof\right) = 1.$$
	The set of \emph{correlated $\epsilon$-equilibria} $\mathcal{Q}_\epsilon$ is the convex
	polytope: [see~\eqref{eq:CE_defn}, shown at the bottom the page], where $\pi^k(i,\actprof^{-k})$ denotes the probability that agent $k$ picks action $i$ and the rest $\actprof^{-k}$. If $\epsilon = 0$, the convex polytope represents the set of correlated equilibria, and is denoted by $\mathcal{Q}$.
\end{definition}
\vspace{0.1cm}

Several reasons motivate adopting the correlated equilibrium in large-scale social networks. It is structurally and computationally simpler than the Nash equilibrium. The coordination among agents in the correlated equilibrium can further lead to potentially higher utilities than if agents take their actions independently (as required by Nash equilibrium)~\cite{Aum87}. Finally, it is more realistic as the observation of the common history of actions naturally correlates agents future decisions~\cite{HM01a}.

An intuitive interpretation of correlated equilibrium is ``coordination in decision-making.'' 
Suppose a mediator is observing a repeated interactive decision making process among multiple selfish agents. The mediator, at each period, gives private recommendations as what action to take to each agent. The recommendations are correlated as the mediator draws them from a joint probability distribution on the action profile of all agents; however, each agent is only given recommendations about her own decision. Each agent can freely interpret the recommendations and decide if to follow. A correlated equilibrium results if neither of agents wants to deviate from the provided recommendation.
That is, in correlated equilibrium, agents' decisions are coordinated as if there exists a global coordinating device that all agents trust to follow.

\section{Regret-Based Collaborative Decision Making}
\label{sec:algorithm}
This section presents the adaptive decision making algorithm that combines the regret-based reinforcement learning procedure~\cite{HM01a}, in the economics literature, with the diffusion cooperation strategies~\cite{Say14,Say14b}, which has recently attracted much attention in the signal processing society. 

\subsection{Agents' Beliefs}
\label{subsec:regret}
Time is discrete $n=1,2,\ldots$. At each time $n$, the agent makes a decision $\actk_\dtimee$ according to a decision strategy $\mixedstrat^\plyrind_\dtimee = (\mixedstratind^\plyrind_\dtimee(1),\cdots,\mixedstratind^\plyrind_\dtimee(A))$ which relies on the agent's belief matrix $\regmatplyr_{\dtimee} = [\regplyrij]$. Each element $\regplyrij$ records the discounted time-averaged regrets---losses in utilities---had the agent selected action $j$ every time it played action $i$ in the past, and is updated via the recursive expression: [see~\eqref{eq:regret} at the bottom of the next page].
%
%
In~\eqref{eq:regret}, $0<\stepsize\ll 1$ is a small parameter that represents the adaptation rate of the strategy update procedure, and is required when agents face a game where the parameters (e.g. utility functions) slowly jump change over time~\cite{NKY13a}. Further, $I(X)$ denotes the indicator operator: $I(X) = 1$ if statement $X$ is true, and 0 otherwise. Note that the update mechanism in~\eqref{eq:regret} relies only on the realized utilities, defined in~\eqref{eq:realized-payoff}.

\begin{figure*}[!hb]
	\hrulefill
	\vspace*{-2pt}
	\setcounter{MYtempeqncnt}{\value{equation}}
	\setcounter{equation}{8}
	\begin{equation}
	\label{eq:strategy-MC-homog-social-group}
	\begin{split}
	\mixedstratind^{\plyrind}_\dtimee(i)= \left\{
	\begin{array}{ll}
	(1-\explor)\min\lbr \frac{1}{\inertia^\plyrind} \big|r^\plyrind_\dtimee\big(\act^\plyrind_{\dtimee-1},i\big)\big|^{+},\frac{1}{\fact}\rbr + \frac{\explor}{\fact}, & i\neq \act^\plyrind_{\dtimee-1}\\
	1-\sum_{j\neq i} p^{\plyrind}_\dtimee(j), & i = \act^\plyrind_{\dtimee-1}
	\end{array}
	\right.
	\end{split}
	\end{equation}
	\setcounter{equation}{\value{MYtempeqncnt}}
	\vspace*{-30pt}
\end{figure*}


Positive $\regplyrij$ implies the opportunity to gain by switching from action $i$ to $j$ in future. Therefore, the regret-matching reinforcement learning procedure, that we present later in this section, assigns positive probabilities to all actions $j$ for which $\regplyrij > 0$. In fact, the probabilities of switching to different actions are proportional to their regrets relative to the current action, hence the name `regret-matching'.

\subsection{Diffusion Cooperation Strategy}
\label{subsec:diffusion}

Inspired by the idea of diffusion least mean squares over adaptive networks~\cite{LS08,Say14b}, we enforce cooperation among neighboring agents via exchanging and fusing regret information. Such diffusion of regret information is rewarding 
since agents belong to the same homophilic group---see Definition~\ref{def:symmetric-game}. That is, all agents attempt to optimize the same utility function, however, in the presence of interdependence among their decisions. It has been shown in~\cite{LS08,Say14b} that such cooperation strategies can lead to faster resolution of uncertainties in decentralized inference and optimization problems over adaptive networks, and enable agents to respond in real time to changes underlying such problem. In view of these benefits, this paper studies, for the first time, application of such diffusion cooperation strategies in the game-theoretic reinforcement learning context in social networks.

At the end of each decision period $n$, 
agent $k$ shares the belief matrix $\regmatplyr_{\dtimee}$ with the neighbors
$\neighborhoodk$ on the network connectivity graph $\G$---see Definition~\ref{def:connectivity-graph}. Agent $k$ then fuses the collected information via a linear combiner~\cite{Say14,Say14b}:
\setcounter{equation}{5}
\begin{equation}
\label{eq:linear-combiner}
\textstyle \bar{R}^k_n := \sum_{l\in\G^k} w_{_{kl}} R^l_n
\end{equation}
where $w_{_{kl}}$ denotes the weight that agent $k$ assigns to the regrets experienced by agent $l$ in her immediate neighborhood on the network connectivity graph. These weights give rise to a global \emph{weight matrix} $W = [w_{_{kl}}]$ in the network of agents. In this paper, we assume:
\begin{equation}
\label{eq:W}
W := I_{K} + \stepsize C,
\end{equation}
where $\stepsize$ is the same as the step-size in~\eqref{eq:regret}, and 
\begin{equation}
\label{eq:C}
\begin{array}{c}
C^\prime = C, \; C\mathbf{1}_{\fplyr} = \mathbf{0}_{\fplyr}, \;\textmd{and}\\
|c_{\plyrind l}|\leq 1,\;c_{\plyrind l}\geq 0 \;\;\textmd{for}\;\; \plyrind\neq l, c_{\plyrind l} > 0 \;\;\textmd{iff} \;\; (\plyrind,l)\in\gameedge.
\end{array}
\end{equation}
Here, $I_K$ denotes the $K\times K$ identity matrix, $(\cdot)^\prime$ denotes the transpose operator,  $\mathbf{1}_{K}$ and $\mathbf{0}_{K}$ represent $K\times 1$ vector of all ones and zeros, respectively. Agent $k$ then combines the fused information with her own realized utility at the current period, and updates the decision policy for the next period.

It is shown in~\cite{NKY13b} that, by properly rescaling the periods at which observation of individual agents and fusion of neighboring beliefs take place, the standard diffusion cooperation strategy in~\cite{LS08} can be approximated by the diffusion strategy with the weight matrix $W$ in~\eqref{eq:W}. This further allows using the well-known ordinary differential equation (ODE) method~\cite{BMP90,KY03} for the convergence analysis. In light of~\eqref{eq:linear-combiner}, the belief of each agent is a function of both her own past experience and those of neighboring agents. This enables them to respond in real-time to the non-stationarities underlying the game.

\subsection{Regret-Matching With Diffusion Cooperation}
\label{subsec:alg}
The proposed reinforcement learning algorithm 
is summarized in the following protocol that mimics human's learning process:

\begin{figure}[!t]
	\setlength{\abovecaptionskip}{0em}
	\setlength{\belowcaptionskip}{0em}
	\begin{center}
	\vspace{-0.3cm}
		\includegraphics[width=0.4\textwidth]{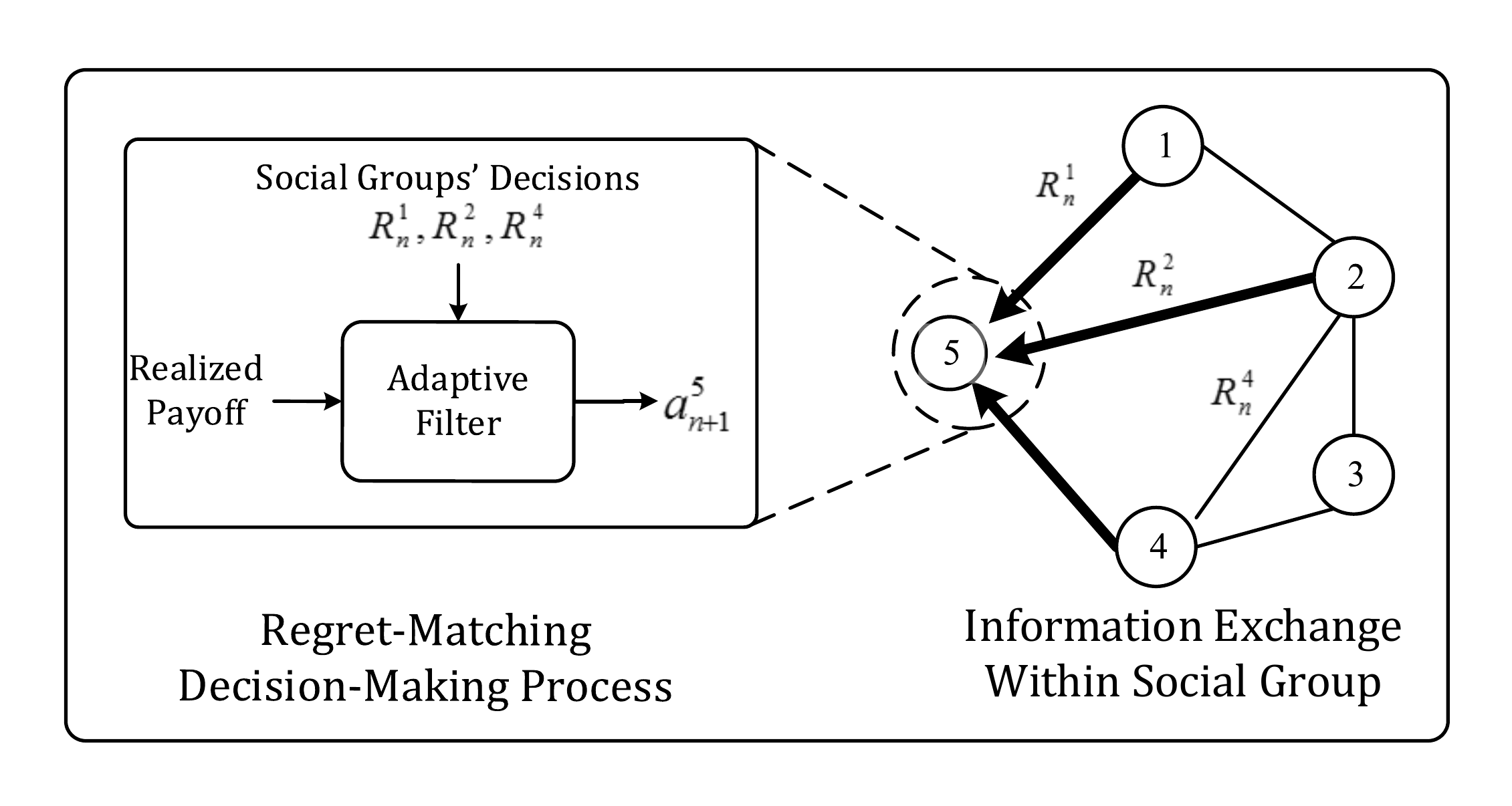}
	\end{center}
	\caption{Regret-matching with diffusion cooperation. 
	}
	\label{fig:game-alg}
\end{figure}

\vspace{0.2cm}
\textbf{Social Decision Protocol}

\vspace{0.1cm}
\begin{enumerate}
	\item[] \textbf{Step 1:} Individual $k$ chooses action $a_n^k$ randomly from a weight vector (probabilities)  $\mixedstrat_n^k$. This weight vector is an ordinal function\footnote{An ordinal function orders pairs of alternatives such that one is considered to be worse than, equal to, or better than the other---see item c) in Sec.~\ref{subse:discussion-alg}} of regret due to its previous actions.
	\vspace{-0.3cm}
	\item[] \textbf{Step 2:} The individual updates regrets based on its actions and the associated outcomes as
	$\textstyle\sum_{\tau=1}^n (1-\stepsize)^{n-\tau} F(a_{\tau}^k),$
	where $F\cd$ is an ordinal function of the action. The exponential discounting places more importance on recent actions.
	\vspace{0.1cm}
	\item[] \textbf{Step 3:} The individual then shares and fuses its regrets with other individuals in the social group.
\end{enumerate}
\vspace{0.2cm}

Below we abstract the above social decision protocol into Algorithm~\ref{alg:homo-social-clique} so as to facilitate analysis of the global behavior; see also Fig.~\ref{fig:game-alg} for a schematic illustration of Algorithm~\ref{alg:homo-social-clique}. 
The first two steps implement the regret-matching reinforcement learning procedure~\cite{HM00}, whereas the last step implements the diffusion protocol~\cite{LS08,Say14b}.

\begin{algorithm}[!t]
	
	\textbf{Initialization}:  Set $$\inertia^\plyrind > A\left|\utilityk_{\max} - \utilityk_{\min} \right|,$$ where $\utilityk_{\max}$ and $\utilityk_{\min}$ denote the upper and lower bounds on the utility function, respectively. Set the step-size $0<\stepsize\ll 1$, and initialize
	$$\mixedstrat^\plyrind_{_0} = (1/ \fact)\cdot\mathbf{1}_{\fact}, \regmatplyr_{_0} = \mathbf{0}.$$
	
	
	\textbf{Step 1:} Choose Action Based on Past Regret.
	$$\act^\plyrind_\dtimee\sim\mixedstrat^\plyrind_\dtimee = \left(\mixedstratind^\plyrind_\dtimee(1),\ldots,\mixedstratind^\plyrind_\dtimee(\fact)\right),$$ 
	where $\mixedstratind^\plyrind_\dtimee(i)$ is given in~\eqref{eq:strategy-MC-homog-social-group}, and $\left|x\right|^{+} = \max\lbrace0,x\rbrace$.
	
	\textbf{Step 2:} Update Individual Regret.
	\setcounter{equation}{9}
	\begin{equation}
	\label{eq:reg-update-4}
	\regmatplyr_{\dtimenext} = \regmatplyrdiff_\dtimee + \stepsize\lb F^\plyrind \left(\act^\plyrind_{\dtimee}\right) - \regmatplyrdiff_\dtimee \rb
	\end{equation}
	where $F^\plyrind \left(\actk_\dtimee\right) = [f_{ij}(\actk_\dtimee)]$ is an $A \times A$ matrix with elements
	\begin{equation}
	\label{eq:F}
	f^{\plyrind}_{ij}\left(\act^\plyrind_{\dtimee}\right) =  \frac{\mixedstratind^\plyrind_\dtimee(i)}{\mixedstratind^\plyrind_\dtimee(j)}\utilityk_\dtimee\big(\act^\plyrind_{\dtimee}\big)\cdot \indicatorj - \utilityk_\dtimee\big(\act^\plyrind_{\dtimee}\big)\cdot\indicatori.\nonumber
	\end{equation}

	\textbf{Step 3:} Fuse Regrets with Members of Social Group.
	\begin{equation}
	\label{eq:linear-combiner-2}
	\textstyle\regmatplyrdiff_{\dtimenext} = \sum_{l\in\cneighborhoodk} \weightkl \bar{R}^l_{\dtimee}.
	\end{equation}
	
	\textbf{Recursion}.
	Set $\dtimee\leftarrow \dtimenext$, and go Step 1.
	\caption{Regret-Matching With Diffusion Cooperation}
	\label{alg:homo-social-clique}
\end{algorithm}

\subsection{Discussion and Intuition}
\label{subse:discussion-alg}
Distinct properties of the local adaptation and learning algorithm summarized in Algorithm~\ref{alg:homo-social-clique} are as follows:

\paragraph{Decision strategy} The randomized strategy~\eqref{eq:strategy-MC-homog-social-group} is simply a weighted average of two probability vectors: The first term, with weight $1 - \explor$, is proportional to the positive part of the regrets. Taking the minimum with $1/\fact$ guarantees that $\mixedstrat^\plyrind_n$ is a valid probability distribution, i.e., $\sum_{i} \mixedstratind^\plyrind_n(i) = 1$. The second term, with weight $\explor$, is just a uniform distribution over the action space~\cite{NKY13a,HM01a}. It forces every action to be played with some minimal frequency (strictly speaking, with probability $\explor/\fact$). The \emph{exploration} factor $\explor$ is essential to be able to estimate the contingent utilities using only the realized utilities; it can, as well, be interpreted as exogenous statistical ``noise.'' As will be discussed later, larger $\explor$ will lead to the convergence of the global behavior to a larger $\CEdistance$-distance of the correlated equilibria set.

\paragraph{Adaptive behavior} In~\eqref{eq:reg-update-4}, $\stepsize$ essentially introduces an exponential forgetting of the experienced regrets in the past, and facilitates adaptivity to the evolution of the non-cooperative game model over time. As agents successively take actions, the effect of the old experiences on their current decisions vanishes. 
This enables tracking time variations on a timescale that is as fast as the adaptation rate of Algorithm~\ref{alg:homo-social-clique}.

\paragraph{Ordinal choice of actions} The decision strategy is an ordinal function of the experienced regrets. Actions are ordered based on the regret values with the exception that all actions with negative regret are considered to be equally desirable---see footnote~\ref{footnote:ordinal}. 

\paragraph{Inertia} The choice of $\inertia^\plyrind$ guarantees that there is always a positive probability of picking the same action as the last period. Therefore, $\inertia^\plyrind$ can be viewed as an ``inertia'' parameter.
It mimics humnas' decision making process and plays a significant role in breaking away from bad cycles. 
This inertia is, in fact, the very factor that makes convergence to the correlated equilibria set possible under (almost) no structural assumptions on the underlying game~\cite{HM00}.



\paragraph{Markov chain construction} The sequence $\lbr\actk_n,\regmatplyr_n\rbr$
is a Markov chain with state space $\actset$, and transition probability matrix
\begin{equation}
P\big(\actk_n = i\big|\actk_{n-1} = j\big) = P_{ji}\big(\regmatplyr_n\big).\nonumber
\end{equation}
The above transition matrix is continuous, irreducible and aperiodic for each $\regmatplyr_n$.
It is (conditionally) independent of other agents' action profile, which may be correlated. More precisely, let $\mathbf{h}_n =\lbr\actprof_{\tau}\rbr_{\tau=1}^n$, where $\actprof_\tau = (\actk_\tau,\actprof^{-\plyrind}_\tau)$
denote the history of decisions made by all agents up to time $n$. Then,
\begin{equation}
\label{eq:independence-actions}
\begin{split}
&\textmd{Pr}\big(\actk_n = i,\actprof^{-k}_n = \bar\actprof\;|\;\mathbf{h}_{n-1}\big)\\
&\quad=\textmd{Pr}\big(\actk_n = i\;\big|\;\mathbf{h}_{n-1}\big)P\big(\actprof^{-\plyrind}_\dtimee = \bar\actprof\;|\;\mathbf{h}_{n-1}\big)\\
&\quad=P_{\actk_{n-1}i}\big(\regmatplyr_\dtimee\big)P\big(\actprof^{-\plyrind}_{\dtimee} = \bar\actprof\;|\;\mathbf{h}_{n-1}\big).
\end{split}
\end{equation}
The sample path of this Markov chain $\{\actk_{\dtimee}\}$ is fed back into the stochastic approximation algorithm that updates $\regmatplyr_{\dtimee}$, which in turn affects its transition matrix. This interpretation is useful in Sec.~\ref{sec:proof-game} when deriving the limit dynamical system representing the behavior of Algorithm~\ref{alg:homo-social-clique}.

\section{Emergence of Rational Global Behavior}
\label{sec:main-results-game}
This section characterizes the global behavior emerging from agents individually following Algorithm~\ref{alg:homo-social-clique}. 

\subsection{Global Behavior}
\label{subsec:global-behavior}
The global behavior $\z_\dtimee$ of the network at each time $n$ is defined as the \emph{discounted empirical frequency} of joint action profile of all agents up to period $\dtimee$. Formally,
\begin{equation}
\label{eq:global-behavior}
\textstyle\z_\dtimee = (1-\stepsize)^{n-1}\e_{\actprof{_1}} + \stepsize\sum_{2\leq\tau\leq k} (1-\stepsize)^{k-\tau} \e_{\actprof_{\tau}},
\end{equation}
where $\e_{\actprof_{\tau}}$ is a unit vector on the space of all possible joint action profiles $\actset^{\plyrset}$ with the element corresponding to the joint play $\actprof_{\tau}$ being equal to one. The small parameter $0<\stepsize\ll 1$ is the same as the adaptation rate in~\eqref{eq:reg-update-4}. It introduces an exponential forgetting of the past decision profiles to enable adaptivity of the network behavior to the evolution of the game model. That is, the effect of the old game model on the decisions of agents vanishes as they repeatedly take actions. Given $\z_{\dtimee}$, the average utility accrued by each agent can be straightforwardly evaluated, hence the name global behavior. It is more convenient to define $\z_\dtimee$ via the stochastic approximation recursion
\begin{equation}
\label{eq:global-behavior-SA}
\z_\dtimee = \z_{n-1} + \stepsize \lb \e_{\actprof_\dtimee} - \z_{n-1}\rb.
\end{equation}
%



\subsection{Asymptotic Local and Global Behavior}
\label{subsec:CE-convergence}
In what follows, we present the main theorem that reveals both the local and global behavior emerging from each agent individually following Algorithm~\ref{alg:homo-social-clique} in a static game model. The regret matrices $\regmatplyr_\dtimee$, $\plyrind\in\plyrset$, and $\z_\dtimee$ will be used as indicatives of agent's local and global experience, respectively.

We use stochastic averaging~\cite{KY03}
in order to characterize the asymptotic behavior of Algorithm~\ref{alg:homo-social-clique}. The basic idea is that, via a `local' analysis, the noise effects in the stochastic algorithm is averaged out so that the asymptotic behavior is determined by that of a `mean' dynamical system.
To this end, in lieu of working with the discrete-time iterates directly, one works with continuous-time interpolations of the iterates. Accordingly, define the piecewise constant interpolated processes
\begin{equation}
\label{eq:interpolated-game}
\regmatplyrinterpol = \regmatplyr_\dtimee, \;\z^{\stepsize}(t) = \z_\dtimee\;\;\mbox{for}\;\; t\in[n\stepsize,(n+1)\stepsize).
\end{equation}
Further, with slight abuse of notation, denote by $\regmatplyr_\dtimee$ the regret matrix rearranged as vectors of length $(A)^2$---rather than an $A\times A$ matrix---and let $\regmatplyrinterpolcdot$ represent the associated interpolated vector processes; see~\eqref{eq:interpolated-game}. Let further $\|\cdot\|$ denote the Euclidean norm, and $\RR_{-}$ represent the negative orthant in the Euclidean space of appropriate dimension. The following theorem characterizes the local and global behavior emergent from following Algorithm~\ref{alg:homo-social-clique}.
\vspace{0.1cm}
\begin{theorem}
	\label{theorem:main-game}
	Let $t_{\stepsize}$ be any sequence of real numbers satisfying $t_{\stepsize}\to\infty$ as $\stepsize\to 0$. For each $\epsilon$, there exists an upper bound $\widehat{\explor}(\epsilon)$ on the exploration parameter $\explor$ such that, if every agent follows Algorithm~\ref{alg:homo-social-clique} with $0<\explor < \widehat{\explor}(\epsilon)$ in~\eqref{eq:strategy-MC-homog-social-group}, as $\stepsize\to 0 $, the following results hold:
	
	(i) The regret vector $R^{\plyrind,\stepsize}(\cdot+t_\stepsize)$ converges in probability to an $\epsilon$-distance of the negative orthant. That is, for any $\beta>0$,
	\begin{equation}
	\label{eq:convergence-1}
	\lim_{\stepsize\to 0} P\left(\mathrm{dist}\big[ R^{\plyrind,\stepsize}(\cdot+t_\stepsize),\RR_{-}\big] - \epsilon > \beta \right) = 0
	\end{equation}
	where $\mathrm{dist}[\cdot,\cdot]$ denotes the usual distance function.
	
	(ii) The global behavior vector $\z^\stepsize(\cdot+t_\stepsize)$ converges in probability to the correlated $\epsilon$-equilibria set $\mathcal{C}_{\CEdistance}$ in the sense that
	\begin{equation}
	\label{eq:convergence-2}
	\mathrm{dist}\big[ \z^{\stepsize}(\cdot+t_{\stepsize}),\mathcal{C}_{\CEdistance}\big] = \inf_{\z \in \mathcal{C}_{\CEdistance}}\left\| \z^{\stepsize}(\cdot+t_{\stepsize}) - \z\right\| \rightarrow 0.
	\end{equation}
	\begin{IEEEproof}
		See Appendix~\ref{sec:proof-game} for a sketch of the proof.
	\end{IEEEproof}
\end{theorem}
\vspace{0.1cm}

The above theorem simply asserts that, if an agent individually follows Algorithm~\ref{alg:homo-social-clique}, she will experience regret of at most $\CEdistance$ after sufficient repeated plays of the game.  Indeed, $\CEdistance$ can be made arbitrarily small by properly choosing the exploration parameter $\explor$ in~\eqref{eq:strategy-MC-homog-social-group}. It further states that, if now all agents in the networked multi-agent system start following Algorithm~\ref{alg:homo-social-clique} independently, their collective behavior converges to the correlated $\epsilon$-equilibria set. Differently put, agents can coordinate their strategies in a distributed fashion so that the distribution of their joint behavior is close to the correlated equilibria polytope. From the game-theoretic point of view, it shows that non-fully rational local behavior of agents---due to utilizing a `better-response' rather than a `best-response' strategy---can lead to the manifestation of globally sophisticated and rational behavior at the network level. Note in the above theorem that the convergence arguments are to a set rather than a particular point in that set. 

\begin{remark}
The constant step-size in Algorithm~\ref{alg:homo-social-clique} enables it to adapt to changes underlying the game model. 
Using weak convergence methods~\cite{KY03}, it can be shown that the first result in Theorem~\ref{theorem:main-game} holds if the parameters underlying the game undergo random changes on a timescale that is no faster than the timescale determined by the adaptation rate of Algorithm~\ref{alg:homo-social-clique}. The second result in Theorem~\ref{theorem:main-game} will further hold if the changes occur on a slower timescale. The reader is referred to~\cite{NKY13a} for further details.
\end{remark}

\subsection{Numerical Example}
\label{sec:example}
The limiting behavior of Algorithm~\ref{alg:homo-social-clique} follows a differential inclusion---see the proof of Theorem~\ref{theorem:main-game} in Appendix~\ref{sec:proof-game}. Differential inclusions are generalizations of ordinary differential equations (ODEs) in which the sample paths belongs to a set; therefore, independent runs lead to different sample paths. This prohibits deriving an analytical rate of convergence for reinforcement learning algorithms of this type.
Here, we resort to Monte Carlo simulations to illustrate and compare the performance of Algorithm~\ref{alg:homo-social-clique}. 

\begin{table}[!t]
	\caption{Agents' Payoffs in a Symmetric Non-Cooperative Game}
	\label{table:example}
	\centering
	\renewcommand{\arraystretch}{1.3}
	\begin{tabular}{ll}
		\begin{tabular}{c|cc}
			\multicolumn{1}{c}{}& $a^{2}=1$ & $a^{2}=2$ \\ \cline{2-3}
			$a^{1}=1$ & \multicolumn{1}{c|}{$(2,2,5)$} & \multicolumn{1}{c|}{$(3,6,4)$} \\ \cline{2-3}
			$a^{1}=2$ & \multicolumn{1}{c|}{$(6,3,4)$} & \multicolumn{1}{c|}{$(4,4,6)$} \\ \cline{2-3}
		\end{tabular}
		&
		\begin{tabular}{cc}
			$a^{2}=1$ & $a^{2}=2$ \\ \cline{1-2}
			\multicolumn{1}{|c|}{$(1,1,3)$} & \multicolumn{1}{c|}{$(1,4,5)$} \\ \cline{1-2}
			\multicolumn{1}{|c|}{$(4,1,0)$} & \multicolumn{1}{c|}{$(6,6,4)$} \\ \cline{1-2}
		\end{tabular} \\
		{}&{}\\[-0.2cm]
		\hspace{2.3cm}$a^{3}=1$ & \hspace{0.9cm}$a^{3}=2$
	\end{tabular}
	\vspace{-0.3cm}
\end{table}

\begin{figure}[!t]
	\setlength{\abovecaptionskip}{0em}
	\setlength{\belowcaptionskip}{0em}
	\begin{center}
		\includegraphics[width=2.7in]{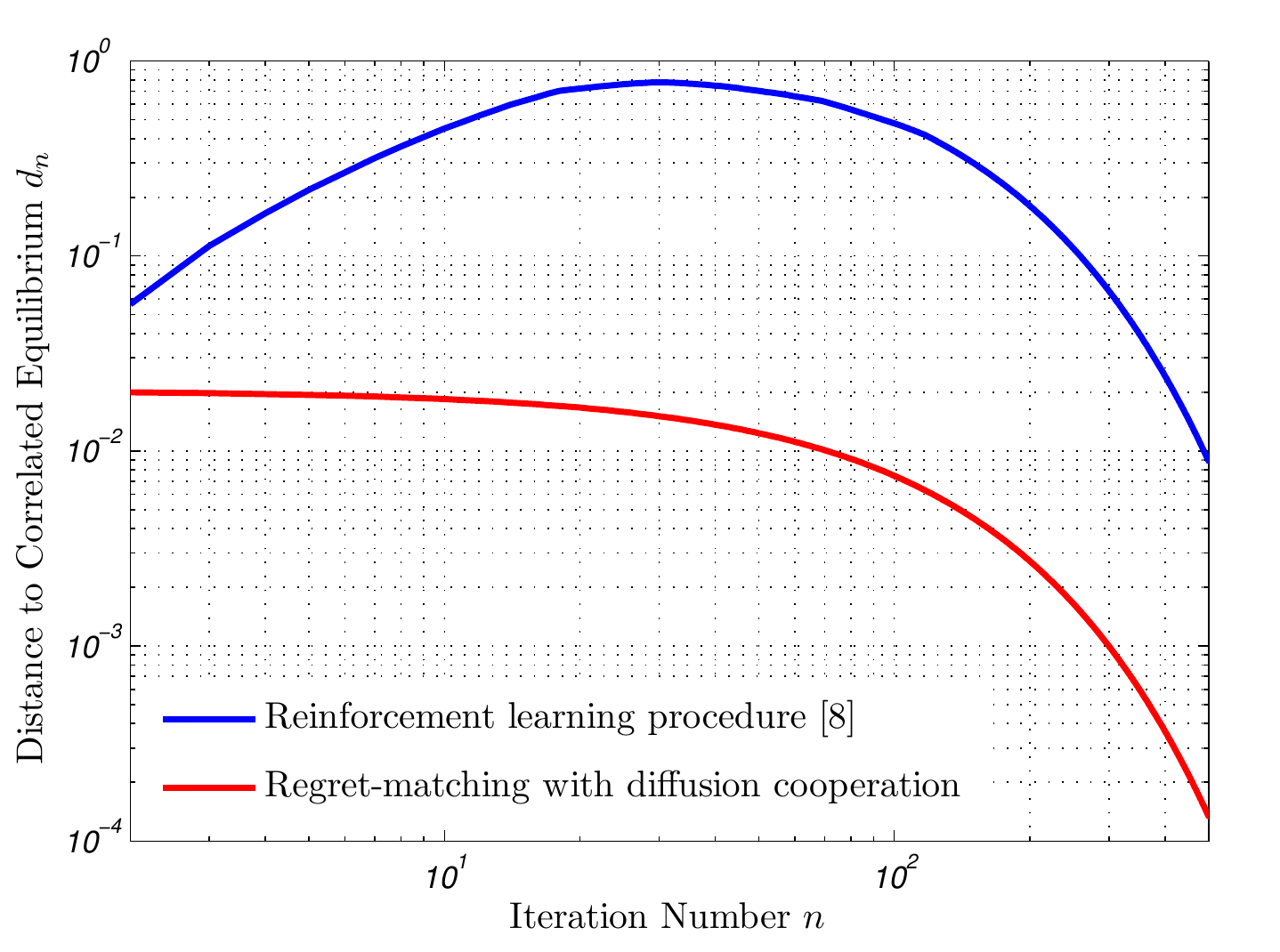}
	\end{center}
	\caption{Distance to correlated equilibrium vs. iteration number.}
	\vspace{-0.2cm}
	\label{fig:game-example}
\end{figure}

Consider a non-cooperative game among three agents $\plyrset = \{1,2,3\}$ with action set $\actset = \{1,2\}$. Agents 1 and 2 exhibit identical homophilic characteristics and, hence, form a social group. That is, $\mathcal{E} = \lbrace(1,2),(2,1)\rbrace$ in the network connectivity graph $\G$---see Definition~\ref{def:connectivity-graph}. In contrast, agent 3 is isolated from agents 1 and 2 and, in fact, unaware of their existence. Table~\ref{table:example} presents agents' utilities in normal form: Each element $(x,y,z)$ in the table represents the utility of agents $1$, $2$, and $3$, respectively, corresponding to the particular choice of action. Note in Table~\ref{table:example} that the game is symmetric between agents 1 and 2. Such situations arise in social networks when a homophilic group of agents aims to coordinate their decisions in response to the actions of other (homophilic groups of) agents.
Further, we set
\begin{displaymath}
C = \left[\begin{matrix} -0.25 & 0.25 \\ 0.25 & -0.25 \end{matrix}\right]
\end{displaymath}
in the weight matrix $W$, defined in~\eqref{eq:W}. That is, agents 1 and 2 place 1/4 weight on the information they receive from their neighbor on the connectivity graph, and 3/4 on their own beliefs. We further set the exploration factor $\explor = 0.15$ in the decision strategy~\eqref{eq:strategy-MC-homog-social-group}, and the step-size $\stepsize=0.01$ in~\eqref{eq:reg-update-4}.

As the benchmark, we use the standard reinforcement learning procedure~\cite{HM01a} to evaluate the performance of Algorithm~\ref{alg:homo-social-clique}. However, we replace its decreasing step-size with the constant step-size $\stepsize$ so as to make the two algorithms both adaptive and comparable. In view of the last step in the proof of Theorem~\ref{theorem:main-game} in Appendix~\ref{sec:proof-game}, the distance to the polytope of correlated equilibrium can be evaluated by the distance of the agents' regrets to the negative orthant. More precisely, we quantify the distance to correlated equilibrium set by
\begin{equation}
d_\dtimee = \max_{k\in\plyrset} \textstyle \sqrt{\smash{\sum_{i,j}}\big(|r^k_n(i,j)|^+\big)^2}.
\end{equation}

Fig.~\ref{fig:game-example} shows how $d_\dtimee$ diminishes with time $\dtimee$ for both algorithms. Each point of the depicted sample paths is an average over 100 independent runs of the algorithms. As is clearly evident, cooperation with neighboring agents over the network topology improves the rate of convergence to the correlated equilibrium. Algorithm~\ref{alg:homo-social-clique} outperforms the reinforcement learning procedure~\cite{HM01a} particularly in the initial stages of the learning process, where sharing experiences (regrets) with neighbors leads to $d_\dtimee$ monotonically decreasing with $\dtimee$. 

\section{Detection of Equilibrium Play in Games} \label{sec:revealed}
\begin{figure}[h!]
  \centering
\includegraphics[width=0.49\textwidth]{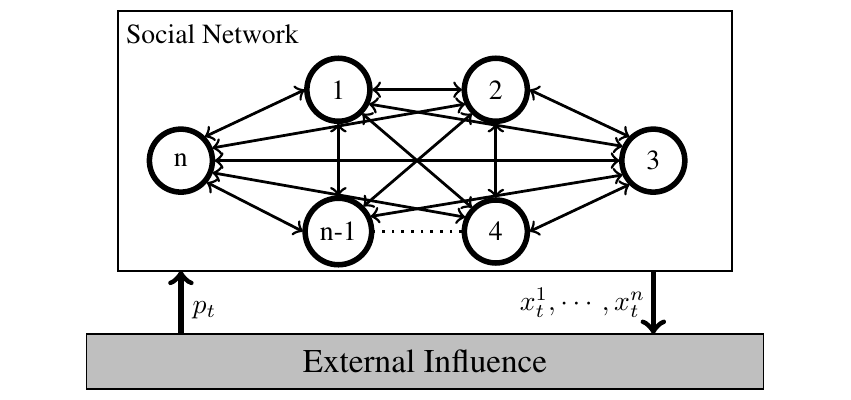}
  \caption{Schematic of a social network containing $\nindx$ interacting agents where $\probe_\tindx\in\mathds{R}^m$  denotes the external influence, and $\response_\tindx^i\in\mathds{R}^m$ the action of agent $i$ in response to the external influence and other agents at time $\tindx$. Note that dotted line denotes consumers $4,\dots,n-1$.  The aim is to determine if the dataset $\dataset=\{(\probe_\tindx,\response_\tindx^1,\response_\tindx^2,\dots,\response_\tindx^\nindx): \tindx\in\{1,2,\dots,\Tindxter\}\}$  is consistent with play from a Nash equilibrium of players engaged in a concave potential game.} 
\label{fig:SNrevealedpreference}
\end{figure}

We now move on to the second part of the paper, namely, using the principle of revealed preferences to detect equilibrium play of agents in a social network. The setup is depicted in Fig.\ \ref{fig:SNrevealedpreference}. The main questions addressed are: Is it possible to detect if the agents are utility maximizers? If yes, can the behavior of the agents be learned using the data from the social network? As mentioned in Sec.~\ref{sec:intro}, these questions are  fundamentally  different to the model-based theme  that is widely used in the signal processing literature in which an 
objective function (typically convex) is proposed and then algorithms are constructed to compute the minimum. In contrast, the revealed preference approach is data-centric--we wish to determine whether the dataset is obtained from  the interaction of utility maximizers. Classical revealed preference theory seeks to determine if an agent is an utility maximizer subject to a budget constraint based on observing its actions over time and is widely studied in the micro-economics literature. The reader is referred to the works \cite{Var12} by Varian (chief economist at Google) for details.

\subsection{Preliminaries: Utility Maximization and Afriat's Theorem}
\label{subsec:afriatstheorem}

Deterministic revealed preference tests for utility maximization were pioneered by Afriat~\cite{Afr67}, and further developed by Diewert~\cite{Die12}, and Varian~\cite{Var83}. Given a time-series of data $\dataset=\{(\probe_\tindx,\response_\tindx), \tindx\in\{1,2,\dots,\Tindxter\}\}$ where $\probe_\tindx\in\reals^m$ denotes the external influence, $\response_\tindx$ denotes the action of an agent, and $\tindx$ denotes the time index, is it possible to detect if the
   agent is an {\it utility maximizer}?
An agent is an {\em utility maximizer} at each time $\tindx$ if for every external influence $\probe_\tindx$, the selected action $\response_\tindx$ satisfies
\begin{equation}
\response_\tindx(\probe_\tindx)\in\operatorname*{arg\,max}_{\{\probe_\tindx^\p \response \leq \budget_\tindx\}}\utility(\response)
\label{eqn:singlemaximization}
\end{equation}
with $\utility(\response)$ a non-satiated utility function. Non-satiated means that an increase in any element of action $\response$ results in the utility function increasing. The non-satiated assumption rules out  trivial cases such as a constant utility function which can be optimized by any action, and as shown by Diewert~\cite{Die12}, without local non-satiation the maximization problem (\ref{eqn:singlemaximization}) may have no solution. In (\ref{eqn:singlemaximization}) the social budget constraint $\probe_\tindx^\p \response_\tindx \leq \budget_\tindx$ denotes the total amount of resources available to the social sensor for selecting the action $\response_\tindx$ in response to the external influence $\probe_\tindx$. An example is the aggregate power consumption of agents in the energy market. The external influence is the cost of using a particular resource, and the action is the amount of resources used. The social impact budget is therefore the total cost of using the resources and is given by $\probe_\tindx^\p \response_\tindx = \budget_\tindx$. Further insight into the social impact budget constraint is provided in Sec.~\ref{sec:ExamplesofEquilibriumPlay}.

The celebrated 
``Afriat's theorem"  provides necessary and sufficient
  conditions for a finite dataset $\dataset$ to have originated from an utility maximizer. 
\begin{theorem}[Afriat's Theorem]{Given a dataset $\mathcal{D}=\{(\probe_t,\response_t):t\in \{1,2,\dots,T\}\}$, the following statements are equivalent:}
\begin{compactenum}
  \item The agent is a utility maximizer and there exists a non-satiated and concave utility function that satisfies (\ref{eqn:singlemaximization}).
\item For scalars $u_t$ and $\lambda_t>0$ the following set of inequalities has a feasible solution:
\begin{equation}
\utility_\tau-\utility_\tindx-\lambda_\tindx \probe_\tindx^\p (\response_\tau-\response_\tindx) \leq 0 \text{ for } \tindx,\tau\in\{1,2,\dots,\Tindxter\}.\
\label{eqn:AfriatFeasibilityTest}
\end{equation}
\item A non-satiated and concave utility function that satisfies (\ref{eqn:singlemaximization}) is given by:
\begin{equation}
\utility(\response) = \underset{\tindx\in T}{\operatorname{min}}\{u_\tindx+\lambda_\tindx \probe_\tindx^\p(\response-\response_\tindx)\}
\label{eqn:estutility}
\end{equation}
  \item The dataset $\mathcal{D}$ satisfies the Generalized Axiom of Revealed Preference (GARP), namely for any $k\leq T$, $\probe_t^\p \response_t \geq \probe_t^\p \response_{t+1} \quad \forall t\leq k-1 \implies \probe_k^\p  \response_k \leq \probe_k^\p  \response_{1}.$ \qed
\end{compactenum}
\label{thrm: Afriat's Theorem}
\end{theorem}

As pointed out in \cite{Var83}, a remarkable feature of Afriat's theorem is that if the dataset can be rationalized by a non-trivial utility function, then it can be rationalized
by a continuous, concave, monotonic utility function. ``Put another way,  violations of continuity, concavity, or monotonicity cannot be detected with only a finite number of demand observations".

Verifying  GARP  (statement 4 of Theorem \ref{thrm: Afriat's Theorem}) on a dataset $\dataset$ comprising $T$ points can be done using Warshall's algorithm with $O(\Tindxter^3)$~\cite{Var83} computations. Alternatively, determining if Afriat's inequalities (\ref{eqn:AfriatFeasibilityTest}) are feasible can be done via a LP feasibility test (using for example interior point methods \cite{BV04}). Note that the utility (\ref{eqn:estutility}) is not unique and is ordinal by construction. Ordinal means that any monotone increasing  transformation of the utility function will also satisfy Afriat's theorem. Therefore the utility mimics the ordinal behavior of humans.
Geometrically the estimated utility (\ref{eqn:estutility}) is the lower envelop of a finite number of hyperplanes that is consistent with the dataset~$\dataset$.

\subsection{Decision Test for Nash Rationality}
\label{subsec:detectiontests}

We now consider a version of Afriat's theorem for deciding if a dataset $\dataset$ from a social network is generated by agents playing from the equilibrium of a potential game. Potential games have been used in telecommunication networking for tasks such as routing, congestion control, power control in wireless networks, and peer-to-peer file sharing~\cite{MT14}, and in social networks to study the diffusion of technologies, advertisements, and influence~\cite{AFPT10}. 

Consider the social network of interconnected agents in Fig.\ \ref{fig:SNrevealedpreference}, given a time-series of data from $n$ agents $\dataset=\{(\probe_\tindx,\response_\tindx^1,\dots,\response_\tindx^n): \tindx\in\{1,2,\dots,\Tindxter\}\}$ with $\probe_\tindx\in\mathds{R}^m$ the external influence, $\response_\tindx^i$ the action of agent $i$, and $\tindx$ the time index, is it possible to detect if the dataset originated from agents that play a potential game? In Fig.\ \ref{fig:SNrevealedpreference} the actions of agents are dependent on both the external influence $\probe_\tindx$ and the actions of the other agents in the social network. The utility function of the agent now includes the actions of other agents--formally if there are $\nindx$ agents, each has a utility function $\utility^i(\response^i,\response_\tindx^{-i})$ with $\response^i$ denoting the action of agent $i$, $\response_\tindx^{-i}$ the actions of the other $\nindx-1$ agents, and $\utility^i(\cdot)$ the utility of agent $i$. Given a dataset $\dataset$, is it possible to detect if the data is consistent with agents playing a game and maximizing their individual utilities? Deb, following Varian's and Afriat's work, shows that refutable restrictions exist for the dataset $\dataset$, given by (\ref{eqn:ResponseData}), to satisfy Nash equilibrium (\ref{eqn:NashEquation})~\cite{Deb09}. These refutable restrictions are however, satisfied by most $\dataset$~\cite{Deb09}. The detection of agents engaged in a concave potential game, and generating actions that satisfy Nash equilibrium, provide stronger restrictions on the dataset $\dataset$~\cite{Deb09}. We denote this behaviour as {\em Nash rationality}, defined as follows:
\begin{definition}[\cite{Deb09}]\label{eq:NashEquilibrium}
Given a dataset
\begin{equation}
 \mathcal{D}=\{(\probe_\tindx,\response_\tindx^1,\response_\tindx^2,\dots,\response_\tindx^\nindx): \tindx\in\{1,2,\dots,\Tindxter\}\},
\label{eqn:ResponseData}
\end{equation}
$\dataset$ is consistent with {\it Nash equilibrium} play if there exist utility functions $\utility^i(\response^i,\response^{-i})$ such that 
\begin{equation}
\response_\tindx^i=\response_\tindx^{i*}(\probe_t)\in\operatorname*{arg\,max}_{\{\probe_\tindx^\p \response^i \leq \budget_\tindx^i\}}\utility^i(\response^i,\response^{-i}). 
\label{eqn:NashEquation}
\end{equation}
In (\ref{eqn:NashEquation}), $\utility^i(\response,\response^{-i})$ is a non-satiated utility function in $\response$, $\response^{-i} = \{\response^j\}_{j\neq i}$ for $i,j\in\{1,2,\dots,\nindx\}$, and the elements of $\probe_\tindx$ are strictly positive. Non-satiated means that for any $\epsilon>0$, there exists a $\response^i$ with $\norm{\response^i-\response^i_\tindx}_2<\epsilon$ such that $\utility^i(\response^i,\response^{-i}) > \utility^i(\response_\tindx^i,\response_\tindx^{-i})$. If  for all $\response^i, \response^j\in \setresponse^i$, there exists a concave potential function $\potfun$ that satisfies
\begin{equation}
\begin{split}
&\utility^i(\response^i,\response^{-i})   -\utility^i(\response^j,\response^{-i}) > 0  \\ 
&\qquad\text{ iff } \potfun(\response^i,\response^{-i})-\potfun(\response^j,\response^{-i}) > 0 
\end{split}
\label{eqn:cordpotfun}
\end{equation} 
for all the utility functions $\utility^i(\cdot)$ with $i\in\{1,2,\dots,\nindx\}$, then the dataset $\dataset$ satisfies {\it Nash rationality}. \qed
 \end{definition}
Just as with the utility maximization budget constraint in (\ref{eqn:singlemaximization}), the budget constraint $\probe_\tindx^\p \response^i \leq \budget_\tindx^i$ in (\ref{eqn:NashEquation}) models the total amount of resources available to the agent for selecting the action $\response^i_\tindx$ to the external influence $\probe_\tindx$.

The following theorem provides necessary and sufficient conditions for a dataset $\dataset$ (\ref{eqn:ResponseData}) to be consistent with Nash rationality (Definition~\ref{eq:NashEquilibrium}). The proof is analogous to Afriat's Theorem when the concave potential function of the game is differentiable~\cite{Deb09,HK14}. 

\begin{theorem}[Multi-agent Afriat's Theorem] Given a dataset $\dataset$ (\ref{eqn:ResponseData}), the following statements are equivalent:
\begin{compactenum}
\item $\dataset$ is consistent with Nash rationality (Definition~\ref{eq:NashEquilibrium}) for an $n$-player concave potential game.
\item Given scalars $v_\tindx$ and $\lambda_\tindx^i>0$ the following set of inequalities have a feasible solution for $\tindx,\tau \in\{1,\dots,T\},$
\begin{equation}
v_{\tau}-v_\tindx-\sum\limits_{i=1}^n\lambda_\tindx^i\probe_\tindx^\p(\response_{\tau}^i-\response_\tindx^i) \leq 0. \quad \label{eqn:NashRationFesTest}
\end{equation}
\item A concave potential function that satisfies (\ref{eqn:NashEquation}) is given by:
\begin{equation}
\hat{V}(\response^1,\response^2,\dots,\response^n) = \underset{t\in T}{\operatorname{min}}\{v_\tindx+\sum_{i=1}^n\lambda_\tindx^ip_\tindx^\p(\response^i-\response^i_\tindx)\}.   \label{egn:est_potential}
\end{equation}
\item The dataset $\mathcal{D}$ satisfies the Potential Generalized Axiom of Revealed Preference (PGARP) if the following two conditions are satisfied. 
\begin{compactenum}
\item For every dataset $\dataset_\tau^i=\{(\probe_\tindx,\response_\tindx^i): \tindx\in\{1,2,\dots,\tau\}\}$ for all $i\in\{1,\dots,n\}$ and all $\tau\in\{1,\dots,T\},$ $\dataset_\tau^i$ satisfies GARP. 
\item The actions $\response_\tindx^i$ originated from agents in a concave potential game. \qed
\end{compactenum}
\end{compactenum}
\label{thrm:NashRationFeasibility}
\end{theorem}
Note that if only a single agent (i.e. $n=1$) is considered, then Theorem~\ref{thrm:NashRationFeasibility} is identical to Afriat's Theorem. Similar to Afriat's Theorem, the constructed concave potential function (\ref{egn:est_potential}) is ordinal--that is, unique up to positive monotone transformations. Therefore several possible options for $\estpotfun(\cdot)$ exist that would produce identical preference relations to the actual potential function $\potfun(\cdot)$. In 4) of Theorem~\ref{thrm:NashRationFeasibility}, the first condition only provides necessary and sufficient conditions for the dataset $\dataset$ to be consistent with a Nash equilibrium of a game, therefore the second condition is required to ensure consistency with the other statements in the Multi-agent Afriat's Theorem. The intuition that connects statements 1 and 3 in Theorem~\ref{thrm:NashRationFeasibility} is provided by the following result from~\cite{MS96}; for any smooth potential game that admits a concave potential function $V$, a sequence of responses $\{\response^i\}_{i\in \{1,2,\dots,n\}}$ are generated by a pure-strategy Nash equilibrium if and only if it is a maximizer of the potential function,
\begin{align}
&\response_\tindx=\{\response_\tindx^1,\response_\tindx^2,\dots,\response_t^n\}\in\operatorname*{arg\,max}V(\{\response^i\}_{i\in \{1,2,\dots,n\}}) \nonumber\\
&\text{s.t. } \quad\probe_\tindx^\p\response^i \leq  I_\tindx^i \quad\quad\forall i\in \{1,2,\dots,n\}
\label{eqn:PotMax}
\end{align}
for each probe vector $\probe_\tindx\in\mathds{R}_+^m$.

The non-parametric test for Nash rationality involves determining if (\ref{eqn:NashRationFesTest}) has a feasible solution. Computing parameters $v_\tindx$ and $\lambda^i_\tindx>0$ in (\ref{eqn:NashRationFesTest}) involves solving a linear program with $T^2$ linear constraints in $(n+1)T$ variables, which has polynomial time complexity~\cite{BV04}. In the special case of one agent, the constraint set in (\ref{eqn:NashRationFesTest}) is the dual of the {\it shortest path problem} in network flows. The parameters $u_t$ and $\lambda_\tindx$ in (\ref{eqn:AfriatFeasibilityTest}) can be computed using  Warshall's algorithm with $O(\Tindxter^3)$~\cite{Var83}.

\subsection{Statistical Test for Nash Rationality}
\label{subsec:FeasibilityTestforNashRationality}

In real world analysis a dataset may fail the Nash rationality test (\ref{eqn:NashRationFesTest}) as a result of the agents actions $\response_\tindx$ being measured in noise. In this section a statistical test is provided to detect for Nash rationality when the actions are measured in noise.

Here we consider additive noise $\noise_\tindx$ such that the measured dataset is given by:
\begin{equation}
\obsdataset=\{(\probe_\tindx,\obsresponse_\tindx^1,\obsresponse_\tindx^2,\dots,\obsresponse_\tindx^n): t\in\{1,2,\dots,T\}\},
\label{eqn:ObservationData}
\end{equation}
consisting of external influences $\probe_\tindx$ and noisy observations of the agents actions $\obsresponse_\tindx^i=\response_\tindx^i+\noise_\tindx^i$. In such cases a feasibility test is required to test if the clean dataset $\dataset$ satisfies Nash rationality (\ref{eqn:NashRationFesTest}). Let $H_0$ and $H_1$ denote the null hypothesis that the clean dataset $\dataset$ satisfies Nash rationality, and the alternative hypothesis that $\dataset$ does not satisfy Nash rationality. In devising a statistical test for $H_0$ vs $H_1$, there are two possible sources of error:
\begin{align}
&\text{\bf Type-I errors:} \text{\hspace{1.6mm}Reject $H_0$ when $H_0$ is valid.} \nonumber\\
&\text{\bf Type-II errors:}  \text{\hspace{0.5mm}Accept $H_0$ when $H_0$ is invalid.}
\label{eqn: hypothesis statements}
\end{align}

Given the noisy dataset $\obsdataset$ (\ref{eqn:ObservationData}) the following statistical test can be used to detect if a group of agents select actions that satisfy Nash equilibrium (\ref{eqn:NashEquation}) when playing a concave potential game:
\begin{equation}
 \boxed{\phantom{\text{\hspace{0.2mm}}}\int\limits_{\Phi^*\{{\bf\obsresponse}\}}^{+\infty}f_M(\psi)\mathrm{d}\psi \overset{H_0}{\underset{H_1}{\gtrless}} \gamma}.
\label{eqn:Statistical_Test}
\end{equation}
In the statistical test (\ref{eqn:Statistical_Test}): \\ (i) $\gamma$ is the ``significance level'' of the statistical test.
\\(ii) The ``test statistic'' $\Phi^*\{{\bf\obsresponse}\}$ is the solution of the following constrained optimization problem for \\ ${\bf\obsresponse}=\{(\obsresponse_\tindx^1,\obsresponse_\tindx^2,\dots,\obsresponse_\tindx^n)\}_{t\in\{1,2,\dots,T\}}$:
\begin{equation}
\begin{array}{rl}
\min & \Phi \\
\mbox{s.t.} & v_{\tau}-v_t-\sum\limits_{i=1}^n\lambda_t^i\probe_t^\p(\obsresponse_{\tau}^i-\obsresponse_t^i)-\sum\limits_{i=1}^n\lambda_t^i\Phi \leq 0 \quad  \\
& \lambda_t^i > 0 \quad \Phi \geq 0 \quad\text{for}\quad t,\tau\in \{1,2,\dots,T\}.\label{eqn:NLP_NashRation}
\end{array}
\end{equation}
(iii) $f_M$ is the probability density function of the random variable $M$ where
\begin{equation}
\textstyle M\equiv\underset{t,\tau}{\text{Max}}\big[\sum_{i=1}^n|\probe_t^\p(\noise_t^i-\noise_{\tau}^i)|\big].
\label{eqn:UpperBoundNLP}
\end{equation}

The following theorem characterizes the performance of the statistical test (\ref{eqn:Statistical_Test}). The proof  is in the appendix.

\begin{theorem} Consider the noisy dataset $\obsdataset$ (\ref{eqn:ObservationData}) of external influences and actions. The probability that the statistical test (\ref{eqn:Statistical_Test}) yields a Type-I error (rejects $H_0$ when it is true) is less then $\gamma$. (Recall $H_0$ and $H_1$ are defined in (\ref{eqn: hypothesis statements})).
\label{thrm:Type-I_ErrorProbability} \qed
\end{theorem}
Note that (\ref{eqn:NLP_NashRation}) is non-convex due to $\sum\lambda_t^i\Phi$; however, since the objective function is given by the scalar $\Phi$, for any fixed value of $\Phi$, (\ref{eqn:NLP_NashRation}) becomes a set of linear inequalities allowing feasibility to be straightforwardly determined~\cite{KH12}.

\subsection{Stochastic Gradient Algorithm to Minimize Type-II Errors}
\label{subsec:SPSA}

Theorem~\ref{thrm:Type-I_ErrorProbability} above guarantees the probability of Type-I errors is less then $\gamma$ for the statistical test (\ref{eqn:Statistical_Test}) for the detection of Nash rationality (Definition~\ref{eq:NashEquilibrium}) from a noisy dataset $\dataset_{obs}$ (\ref{eqn:ObservationData}). In this section, the statistical test (\ref{eqn:Statistical_Test}) is enhanced by adaptively optimizing the external influence vectors ${\bf p}=[\probe_1,\probe_2,\dots,\probe_T]$ to reduce Type-II errors. 

Reducing the Type-II error probability can be achieved by dynamically optimizing the external influence ${\bf \probe}=[\probe_1,\dots,\probe_T]$. The external influence $\bf{\probe}$ is selected as the solution of
{\normalsize 
\begin{align}
&{\bf \probe}^* \in\operatorname*{arg\,min}_{{\bf \probe}\in\mathds{R}^{m\times T}_+}J({\bf \probe}) \nonumber\\
&\quad\quad=\underbrace{\mathds{P}\Bigg(\int\limits_{\Phi^*({\bf y})}^{+\infty}f_{M}(\beta)\mathrm{d}\beta > \alpha \big| \{{\bf \probe, \response(\probe)}\}\in \mathcal{A}\Bigg)}_\text{Probability of Type-II error}. 
\label{eqn:SPSAObjective}
\end{align}
}%
In (\ref{eqn:SPSAObjective}), ${\bf y = \response(\probe)+w}$ with $\bf y$ defined above (\ref{eqn:NLP_NashRation}), $f_M$ is the probability density function of the random variable $M$ (\ref{eqn:UpperBoundNLP}), $\mathds{P}(\cdots|\cdot)$ denotes the conditional probability that (\ref{eqn:Statistical_Test}) accepts $H_0$ for all agents given that $H_0$ is false. The set $\mathcal{A}$ contains all elements $\{{\bf \probe}, {\bf \response(\probe)}\}$, with ${\bf \response(\probe)} = [\response_\tindx^i(\probe_\tindx),\dots,\response_\tindx^n(\probe_\tindx)]$, where $\{{\bf \probe}, {\bf \response(\probe)}\}$ does not satisfy Nash rationality (Definition~\ref{eq:NashEquilibrium}). 

To compute (\ref{eqn:SPSAObjective}) requires a stochastic optimization algorithm as the probability density functions $f_{M}$ are not known explicitly. Given that we must estimate the gradient of the objective function in (\ref{eqn:SPSAObjective}), and that ${\bf \probe}\in\mathds{R}^{m\times T}_+$ can comprise a large dimensional matrix, the simultaneous perturbation stochastic gradient (SPSA) algorithm is utilized to estimate $\bf \probe$ from (\ref{eqn:SPSAObjective})~\cite{Spa03}. The SPSA allows the gradient to be estimated using only two measurements of the objective function corrupted by noise, and for decreasing step size the algorithm with probability one reaches a local stationary point. The SPSA algorithm used to compute $\bf \probe$ is provided below. 

\begin{description}
  \item[\bf Step 1:]\hspace{1.3mm}Choose initial probe ${\bf \probe}_o=[\probe_1,\probe_2,\dots,\probe_T]\in\mathds{R}^{m\times T}_+$
  \item[\bf Step 2:]\hspace{1.3mm}For iterations $q=1,2,3,\dots$
\end{description}
\begin{description}
\item[\hspace{2mm}$\cdot$] Estimate the cost (i.e. probability of Type-II errors) in (\ref{eqn:SPSAObjective}) using 
\begin{equation}
\hat{J}_q({\bf \probe}_q) = \frac{1}{K}\sum\limits_{k=1}^{K}{\bf I}\big(F_M(\Phi^*({\bf y}_k)) \leq 1-\alpha\big)
\label{eqn: SPSA Cost}
\end{equation}
where $\mathbf{I}$ denotes the indicator function, and $F_M(\cdot)$ is an estimate of the cumulative distribution function of $M$ constructing by generating random samples according to (\ref{eqn:UpperBoundNLP}). In (\ref{eqn: SPSA Cost}) $\Phi^*({\bf y}_k)$ is computing using (\ref{eqn:NLP_NashRation}) with the noisy observations ${\bf y}_k = {\bf \response}({\bf \probe}_q)+{\bf w}_k$. Note that ${\bf w}_k$ is a fixed realization of ${\bf w}$, and the dataset $\{{\bf \probe}_q, {\bf \response}({\bf \probe}_q)\}\in\mathcal{A}$ defined below (\ref{eqn:SPSAObjective}). The parameter $K$ in (\ref{eqn: SPSA Cost}) controls the accuracy of the empirical probability of Type-II errors (\ref{eqn:SPSAObjective}). 
\item[\hspace{2mm}$\cdot$] Compute the gradient estimate $\hat{\nabla}_{\bf \probe}\hat{J}_q({\bf \probe}_q)$:
\begin{align}
\hat{\nabla}_{\bf \probe}\hat{J}_q({\bf \probe}_q) &= \frac{\hat{J}_q({\bf \probe}_q+\Delta_q\sigma)-\hat{J}_q({\bf \probe}_q-\Delta_q\sigma)}{2\sigma\Delta_q} \label{eqn: SPSA}\\
\Delta_q(i) &= \begin{cases}
   -1 & \text{with probability 0.5} \\
   +1       & \text{with probability 0.5}
  \end{cases} \nonumber
\end{align}
with gradient step size $\sigma > 0$. 
\item[\hspace{2mm}$\cdot$] Update the probe vector ${\bf \probe}_k$ with step size $\epsilon>0$:
\begin{equation*}
{\bf \probe}_{q+1} = {\bf \probe}_q-\epsilon\hat{\nabla}_{\bf \probe}\hat{J}_q({\bf \probe}_q).
\end{equation*}
\end{description}
The benefit of using the SPSA algorithm is that the estimated gradient $\nabla_{\bf \probe} J_q({\bf \probe}_q)$ in (\ref{eqn: SPSA}) can be computed using only two measurements of the function (\ref{eqn: SPSA Cost}) per iteration; see \cite{Spa03} for the convergence and tutorial exposition of the SPSA algorithm.  In particular, for constant step size $\epsilon$, it converges weakly (in probability) to a local stationary point~\cite{KH12}.

\section{Examples of Equilibrium Play: Energy Market and Detection Malicious Agents}
\label{sec:ExamplesofEquilibriumPlay}

In this section we provide two examples of how the decision test (\ref{eqn:NashRationFesTest}), statistical detection test (\ref{eqn:Statistical_Test}), and stochastic optimization algorithm (\ref{eqn: SPSA}) from Sec.~\ref{sec:revealed} can be applied to detect for Nash rationality (Definition~\ref{eq:NashEquilibrium}) in a social network. The first example uses real-world aggregate power consumption data from the Ontario energy market social network. The second is the detection of malicious agents in an online social network comprised of normal agents, malicious agents, and an authentication agent. 

\subsection{Nash Rationality in Ontario Electrical Energy Market}
\label{subsec:exampleontario}

\begin{figure}[h]
  \centering
\includegraphics[width=0.45\textwidth]{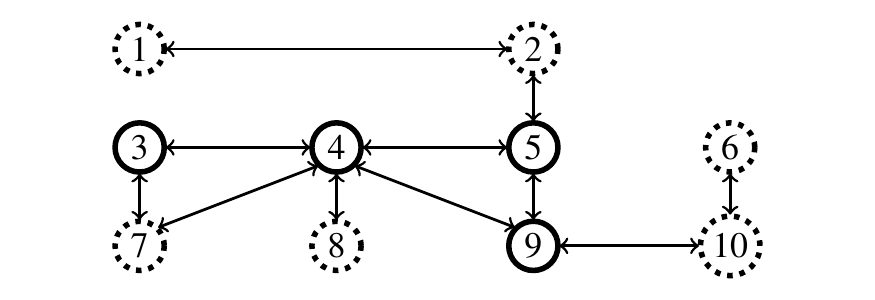}
  \caption{Schematic of the electrical distribution network in the Ontario power grid. The nodes $1,2,\dots,10$ correspond to the distribution zones: Northwest, Northeast, Bruce, Southwest, Essa, Ottawa, West, Niagara, Toronto, and East. The dotted circles indicate zones with external interconnections--these nodes can import/export power to the external network which includes Manitoba, Quebec, Michigan, and New York. The network can be considered a corporate social network of chief financial officers.} 
\label{fig:ontarionetwork}
\end{figure}

In this section we consider the aggregate power consumption of different zones in the Ontario power grid.  A sampling period of $\Tindxter=79$ days starting from January 2014 is used to generate the dataset $\dataset$ for the analysis. All price and power consumption data is available from the {\it Independent Electricity System Operator}\footnote{http://ieso-public.sharepoint.com/} (IESO) website. Each zone is considered as an agent in the corporate network illustrated in Fig.\ \ref{fig:ontarionetwork}. The study of corporate  social networks was pioneered by Granovetter~\cite{Granovetter05} which shows that the social structure of the network can have important economic outcomes. Examples include agents choice of alliance partners, assumption of rational behavior, self interest behavior, and the learning of other agents behavior. Here we test for rational behavior (i.e. utility maximization and Nash rationality), and if true then learn the associated behavior of the zones. This analysis provides useful information for constructing demand side management (DSM) strategies for controlling power consumption in the electricity market. 

The zones power consumption is regulated by the associated price of electricity set by the senior management officer in each respective zone. Since there is a finite amount of power in the grid, each officer must communicate with other officers in the network to set the price of electricity. Here we utilize the aggregate power consumption from each of the $\nindx=10$ zones in the Ontario power grid and apply the non-parametric tests for utility maximization (\ref{eqn:AfriatFeasibilityTest}) and Nash rationality (\ref{eqn:NashRationFesTest}) to detect if the zones are demand responsive. If the utility maximization or Nash rationality tests are satisfied, then the power consumption behaviour is modelled by constructing the associated utility function (\ref{eqn:estutility}) or concave potential function of the game (\ref{egn:est_potential}).

To perform the analysis the external influence $\probe_\tindx$ and action of agents $\response_\tindx$ must be defined. In the Ontario power grid the wholesale price of electricity is dependent on several factors such as consumer behaviour, weather, and economic conditions. Therefore the external influence is defined as $\probe_\tindx=[\probe_\tindx(1),\probe_\tindx(2)]$  with $\probe_\tindx(1)$ the average electricity price between midnight and noon, and $\probe_\tindx(2)$ as the average between noon and midnight with $\tindx$ denoting day. The action of each zone correspond to the total aggregate power consumption in each respective tie associated with $\probe_\tindx(1)$ and $\probe_\tindx(2)$ and is given by $\response_\tindx^i=[\response_\tindx^i(1), \response_\tindx^i(2)]$ with $i\in\{1,2,\dots,\nindx\}$. The budget $I_t^i$ of each zone has units of dollars as $\probe_\tindx$ has units of \$/kWh and $\response_\tindx^i$ units of kWh.

\begin{figure}[h]
  \centering
\includegraphics[width=0.45\textwidth]{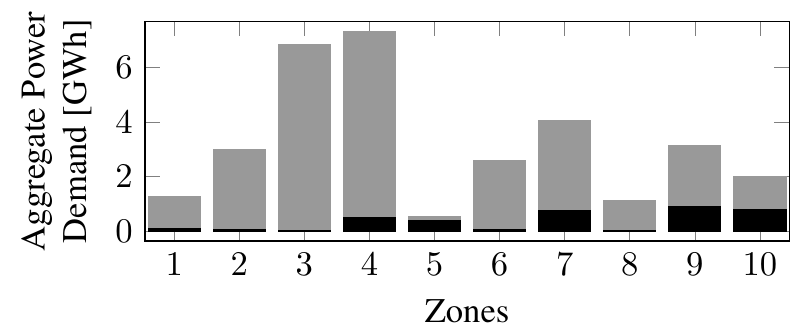}
\vspace{-10pt}
  \caption{Average consumption (gray) and associated noise level $\sigma$ (black) for the price and demand data to satisfy utility maximization in each of the 1,\dots,10 zones in the Ontario power grid defined in Fig.\ \ref{fig:ontarionetwork}. The average hourly consumption over the $T=79$ days starting from January 2014.} 
\label{fig:noiseutilitymaximization}
\end{figure}

We found that the aggregate consumption data of each zone does not satisfy utility maximization (\ref{eqn:AfriatFeasibilityTest}). Is this a result of measurement noise? Assuming the power consumption of agents are independent and identically distributed, the central limit theorem suggests that the aggregate consumption of regions follows a zero mean normal distribution with variance $\sigma^2$. The noise term ${\mathbf w}$ in (\ref{eqn:UpperBoundNLP}) is given by the normal distribution $\mathcal{N}(0,\sigma^2)$. Therefore, to test if the failure is a result of noise, the statistical test (\ref{eqn:Statistical_Test}) is applied for each region, and the noise level $\sigma^2$ estimated for the dataset $\mathcal{D}_{obs}$ to satisfy the $\gamma = 95\%$ confidence interval for utility maximization. The results are provided in Fig.\ \ref{fig:noiseutilitymaximization}. 
As seen from Fig.\ \ref{fig:noiseutilitymaximization}, the Essa, West, Toronto, and East zones do not satisfy the utility maximization requirement. This results as the required noise level $\sigma$ for the stochastic utility maximization test to pass is too high compared with the average power consumption. Therefore if each zone is independently maximizing then only 60\% of the Ontario power grid satisfies the utility maximization test. However it is likely that the zones are engaged in a concave potential game--this would not be a surprising result as network congestion games have been shown to reduce peak power demand in distributed demand management schemes~\cite{ING10}. To test if the dataset $\dataset$ is consistent with Nash rationality the detection test (\ref{eqn:NashRationFesTest}) is applied. The dataset for the power consumption in the Ontario power gird is consistent with Nash rationality. Using (\ref{eqn:NashRationFesTest}) and (\ref{egn:est_potential}), a concave potential function for the game is constructed. Using the constructed potential function, when do agents prefer to consume power? The {\it marginal rate of substitution}\footnote{The amount of one good that an agent is willing to give up in exchange for another good while maintaining the same level of utility.} (MRS) can be used to determine the preferred time for power usage. Formally, the MRS of $\response^i(1)$ for $\response^i(2)$ is given by
\begin{equation*}
 \operatorname{MRS}_{12}=\frac{\partial\hat{V}/\partial \response^i(1)}{\partial\hat{V}/\partial \response^i(2)}.
 \end{equation*} 
From the constructed potential function we find that $\operatorname{MRS}_{12} > 1$ suggesting that the agents prefer to use power in the time period associated with $\response_\tindx(1)$--that is, the agents are willing to give up $\operatorname{MRS}_{12}$ kWh of power in the time period associated with $\response^i(2)$ for 1 additional kWh of power in time period associated with $\response^i(1)$.  

The analysis in this section suggests that the power consumption behavior of agents is consistent with players engaged in a concave potential game. Using the Multi-agent Afriat's Theorem the agents preference for using power was estimated. This information can be used to improve the DSM strategies 
to control power consumption in the electricity market. 

\subsection{Detecting Malicious Agents in Online Social Networks}
\label{subsec:examplemal}
\begin{figure}[h]
  \centering
\includegraphics[width=0.49\textwidth]{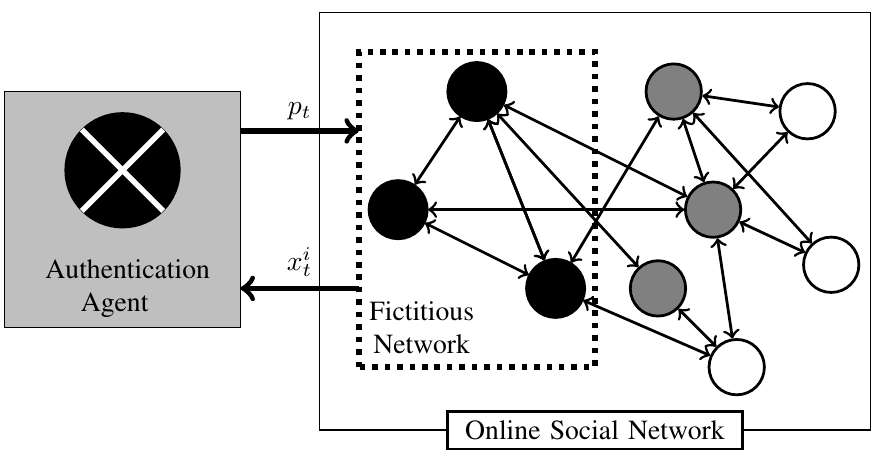}
  \caption{Schematic of an online social network with a network authentication agent that is able to create a fictitious agents (black) to interact with real agents in the social network. Two types of agents are considered: normal agents (white), and malicious agents (grey). The goal is for the authentication agent to be able to detect and eliminate malicious agents from the online social network. The parameters $\probe_\tindx$ and actions $\response_\tindx^i$ are defined in Sec.~\ref{subsec:examplemal}.} 
\label{fig:OSNmalicousdetect}
\end{figure}

Socialbots and spambots are autonomous programs which attempt to
imitate human behavior and are prevalent on popular social networking sites such as Facebook and Twitter. In this section we consider the detection of malicious agents in an online social network comprised of normal agents, malicious agents, and a network authentication agent as depicted in Fig.\ \ref{fig:OSNmalicousdetect}. 

Recent techniques for detecting malicious agents in the social network (i.e. socialbots and spambots) use a method known as {\it behavioural blacklisting} which attempts to detect emails, tweets, friend and follower requests, and URLs which have originated from malicious agents~\cite{RFV07}. Behavioural blacklisting works as socialbots and spambots tend to have different behaviors then humans. For example in Twitter, socialbots and spambots tend to re-tweet far more then normal (i.e. human) agents, and by contrast normal accounts tend to receive more replies, mentions, and re-tweets~\cite{FVDMF14}. The goal of malicious agents is to increase their connectivity in the social network to deliver harmful content such as viruses, gaining followers and friends, marketing, and political campaigning. Consider the network topology depicted in Fig.\ \ref{fig:OSNmalicousdetect}. The authentication agent is designed to detect and eliminate malicious agents in the network. To this end the authentication agent is able to construct fictitious accounts to study the actions of other agents in the network. Denoting $\probe_\tindx\in\mathds{R}_+^m$ as queries for authentication for the $m$ fictitious accounts produced by the authentication node at time $t$, the response of agent $i$ is given by $\response_\tindx^i\in\mathds{R}_+^m$ and is the total number of successfully targeted followers and friends of each of the $m$ fictitious accounts. Note that larger values of $\probe_\tindx$ indicate a stronger quarry for authentication. We consider the following utility function for malicious agents: 
\begin{equation}
\begin{split}
r^i(\vecx^i,\response^{-i}) &= \ln\bigg[\frac{\vecx^i(1)\vecx^i(2)}{\textstyle\sum_{i=1}^n\sum_{j=1}^n\vecx^i(1)\vecx^j(2)}\bigg]\\
 s^i(\vecx^i;\beta) &\textstyle= \ln\Big[\prod_{j=1}^l\Big(1+\frac{\vecx^i(j)}{\beta(j)}\Big)\Big]\\
u^i(\vecx^i,\response^{-i}) &= r^i(\vecx^i,\response^{-i})+s^i(\vecx^i;\beta)
\end{split}
\label{eqn:Agent}
\end{equation}
where $\response_\tindx^{-i}\in\mathds{R}_+^{m\times (n-1)}$ is the actions of the other $(n-1)$ agents, $r$ represents the interdependence of the total targets, and $s$ represents each agents preference to avoid detection. The static inaccuracy (i.e. noise-to-signal ratio) of each quarry authentication is contained in the elements of $\beta\in\mathds{R}_+^m$. The malicious social budget of each agent $i$ is given by $I_\tindx^i$. The total resources available to the authentication agent are limited such that in any operating period $t$ the total resources available for queries for authentication is given by $\sum_{j=1}^{m}\probe_\tindx(j)$. Consider the case with $m=2$ fictitious agents. As the authentication agent commits larger resources to increase the queries for authentication $\probe_\tindx(1)$, the associated number of friends and followers captured by the malicious agent $\response_\tindx(1)$ decreases. Given that the total resources available to the authentication agent is limited, as $\probe_\tindx(1)$ increases $\probe_\tindx(2)$ must decrease. This causes an increase in the friends and followers captured by the malicious agent for the $m=2$ fictitious agent $\response_\tindx(2)$. Therefore the malicious social budget is considered to satisfy the linear relation $I_\tindx^i=\probe_t^\p\response_\tindx^i$. {\it Malicious agents} are those that are engaged in a concave potential game which attempt to maximize their respective utility function (\ref{eqn:Agent}), and {\it normal agents} which have no target preference and therefore select $\response_\tindx^i$ in a uniform random fashion. At each observation $t$, a noisy measurement $\obsresponse_\tindx^i$ (defined in (\ref{eqn:ObservationData})) is made of the actions $\response_\tindx^i$. Given the dataset $\dataset_{obs}$ (\ref{eqn:ObservationData}), a statistical test can be used to detect if malicious agents are present.

The dataset $\dataset$ (\ref{eqn:ResponseData}) for malicious agents are generated by computing the maximum, $\{\response_\tindx^i\}_{i\in\{1,2,\dots,n\}}$, of the concave potential function $V=\sum_{i=1}^n u^i(\response^i,\response^{-i})$, with $u^i(\cdot)$ defined by (\ref{eqn:Agent}), for a given probe $\probe_\tindx$ (refer to (\ref{eqn:PotMax})). The parameter values for the numerical example are $n=3, m=2, \beta=[0.03,0.08]$, and $\gamma=0.05$, where $n,m,\beta$ are defined in Sec.~\ref{subsec:examplemal}. The malicious social budget for each of the $n=3$ agents is generated from the normal distributions: $I_\tindx^1\sim\mathcal{N}(20,1)$, $I_\tindx^2\sim\mathcal{N}(50,1)$, and $I_\tindx^2\sim\mathcal{N}(80,4)$. The queries for authentication are generated from the uniform distribution $\probe_\tindx\sim\mathcal{U}(1,5)$. For normal agents, $\dataset$ (\ref{eqn:ResponseData}) is constructed from $\response_\tindx^i$ obtained from the uniform random variable $\response_\tindx^i\sim\mathcal{U}(1,50)$. The datasets $\dataset_{obs}$ (\ref{eqn:ObservationData}) are obtained using the clean dataset $\dataset$, and additive noise $\noise^i~\sim\mathcal{U}(0,\kappa)$ where $\kappa$ represents the magnitude of the measurement error. 

\begin{figure}[h]
  \centering
\includegraphics[width=0.45\textwidth]{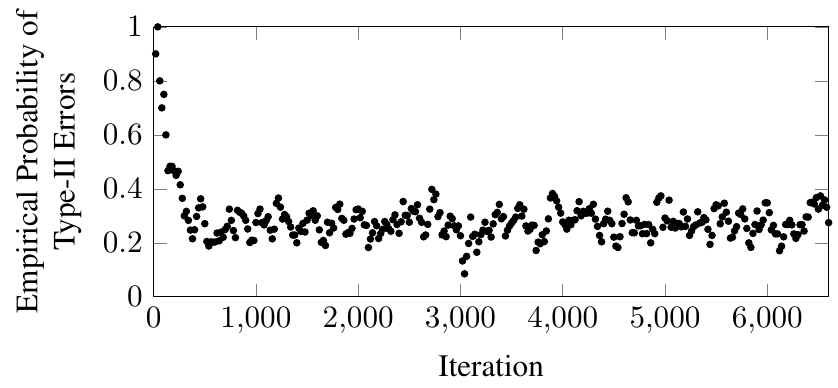}
\vspace{-10pt}
  \caption{Performance of the SPSA algorithm (\ref{eqn: SPSA}) for computing the locally optimal external influence ${\bf \probe}$ to reduce the probability of Type-II errors of the statistical test (\ref{eqn:Statistical_Test}). The parameters are defined in Sec.~\ref{subsec:examplemal}.}
\label{fig:SPSA_performance}
\end{figure}

Fig.\ \ref{fig:SPSA_performance} plots the estimated cost (\ref{eqn: SPSA Cost}) versus iterates generated by the SPSA algorithm (\ref{eqn: SPSA}) for $\sigma=0.1, \epsilon=0.2, \kappa=0.1,\text{ and }T=20$ observations. Fig.\ \ref{fig:SPSA_performance} illustrates that by judiciously adapting the external influence via a stochastic gradient algorithm, the probability of Type-II errors can be decreased to approximately 30\% allowing the statistical test (\ref{eqn:Statistical_Test}) to adequately reject normal agents. 

\begin{figure}[h]
  \centering
\includegraphics[width=0.45\textwidth]{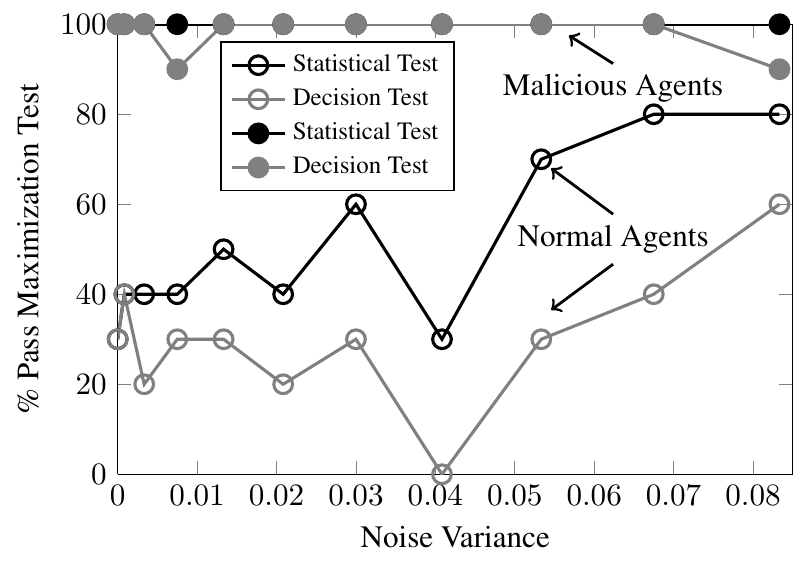}
\vspace{-10pt}
  \caption{Performance of the decision test (\ref{eqn:NashRationFesTest}), and statistical test (\ref{eqn:Statistical_Test}) for the detection of malicious agents and normal agents. The parameters are defined in Sec.~\ref{subsec:examplemal}.}
\label{fig:StatTest_performance}
\end{figure}

Fig.\ \ref{fig:StatTest_performance} plots the probability that a dataset $\dataset_{obs}$ (\ref{eqn:ObservationData}) that satisfies the decision test (\ref{eqn:NashRationFesTest}) and statistical test (\ref{eqn:Statistical_Test}) for agents engaged in a concave potential game. The locally optimized external influence ${\bf \probe}$ was obtained from the results of the SPSA algorithm above, allowing the malicious and normal agents to be distinguished. As seen, the occurrence of Type-I errors in the statistical test is less then $5\%$, as expected from Theorem~\ref{thrm:Type-I_ErrorProbability} because $\gamma=5\%$.

\section{Summary}
\label{sec:conclusion}
The unifying theme of this paper was to study equilibrium play in non-cooperative games amongst agents in a social network. 
The first part focused on the distributed reinforcement learning aspect of an equilibrium notion, namely, correlated equilibrium. Agents with identical homophilic characteristics formed social groups wherein they shared past experiences over the network topology graph. A reinforcement learning algorithm was presented that, relying on diffusion cooperation strategies in adaptive networks, facilitated the learning dynamics. It was shown that, if all agents follow the proposed algorithm, their global behavior of the network of agents is attracted to the correlated equilibria set of the game. 
The second part focused on parsing datasets from a social network for detecting play from the equilibrium of a concave potential game. A non-parametric decision test and statistical test was constructed to detect equilibrium play which only required the external influence and actions of agents. 
To reduce the probability of Type-II errors, a stochastic gradient algorithm was given to adapt the external influence in real time. Finally, we illustrated the application of the decision test, statistical test, and stochastic gradient algorithm in a real-world example using the energy market, and provided a numerical example to detect malicious agents in an online social network. An important property of both aspects considered in this paper is their ordinal nature, which provides a useful approximation to human behavior.


%


\appendices
\section{Sketch of the Proof of Theorem~\ref{theorem:main-game}}
\label{sec:proof-game}
The convergence analysis is based on~\cite{BHS06} and is organized into three steps. For brevity and better readability, details for each step of the proof are omitted, however, adequate references are provided for the interested reader. 

\subsubsection*{Step~1} The first step uses weak convergence methods to characterize the limit individual behavior of agents following Algorithm~\ref{alg:homo-social-clique} as a dynamical system represented by a differential inclusion. Differential inclusions are generalizations of the ODEs~\cite{AC84}. Below, we provide a precise definition.

\begin{definition}
\label{def:diff-inclusion}
A differential inclusion \index{differential inclusion} is a dynamical system of the form
\begin{equation}
\label{eq:F_x}
\frac{d}{dt}X\in \mathcal{F}\left(X\right)
\end{equation}
where $X\in\mathbb{R}^r$ and $\mathcal{F}:\mathbb{R}^r\rightarrow\mathbb{R}^r$ is a Marchaud map~\cite{AC84}. 
That is, i) the graph and domain of $\mathcal{F}$ are nonempty and closed; ii) the values $\mathcal{F}\left(X\right)$ are convex; and iii) the growth of $\mathcal{F}$ is linear: There exists $C >0$ such that, for every $X\in\mathbb{R}^r$,
\begin{equation}
\label{eq:linear_growth}
\sup_{Y\in \mathcal{F}\left(X\right)} \left\| Y\right\|\leq C \left(1 + \left\|X\right\|\right)
\end{equation}
where $\|\cdot\|$ denotes any norm on $\mathbb{R}^r$.
\end{definition}

We proceed to study the properties of the sequence 
$\{\actk_n\}$ made according to Algorithm~\ref{alg:homo-social-clique}, which forms finite-state Markov chain---see Sec.~\ref{subse:discussion-alg}. Standard results on Markov chains show that the transition matrix~\eqref{eq:strategy-MC-homog-social-group} admits (at least) one invariant measure, denoted by $\boldsymbol{\sigma}^{\plyrind}$. Then, the following lemma characterizes the properties of such an invariant measure.
\begin{lemma}
\label{lemma:stationary-distr}
The invariant measure $\boldsymbol{\sigma}^{\plyrind}(\regmatplyr)$ of the transition probabilities~\eqref{eq:strategy-MC-homog-social-group} takes the form
\begin{equation}
\label{eq:statdistr-groups}
\boldsymbol{\sigma}(\regmatplyr) = (1-\explor)\statdistk\big(\regmatplyr\big) + \big(\explor/\fact\big)\cdot\mathbf{1}_{\fact},
\end{equation}
where $\statdistk(\regmatplyr)$ satisfies
\begin{equation}
\label{eq:stat-distr-prop}
\textstyle \sum_{j\neq i} \statdistindj  \big| r^\plyrind(j,i)\big|^+ = \statdistindi  \sum_{j\neq i}  \big| r^\plyrind(i,j) \big|^+,
\end{equation}
and $|x|^+ = \max\{x,0\}$.
\end{lemma}

In light of the diffusion protocol~\eqref{eq:linear-combiner}, agents' successive decisions affect, not only their own future decision strategies, but also their neighbors' policies. This suggests looking at the dynamics of the regret for the entire network: $\globalregret_\dtimee := \textmd{col}\left( R^{1}_\dtimee,\ldots,R^{\fplyr}_\dtimee\right)$.
%
%
Using techniques from the theory of stochastic approximations~\cite{KY03}, we work with the piecewise constant continuous-time interpolations of $\globalregret_\dtimee$, defined by
\begin{equation}
\label{eq:global-regret-interpol}
\globalregretinterpol\ctimeet = \globalregret_\dtimee\;\;\textmd{for}\;\; t\in[k\stepsize,(k+1)\stepsize),
\end{equation}
to derive the limiting process associated with $\globalregret_\dtimee$.
Let $\Delta \actset^{-k}$ represent the simplex of all probability distributions over the joint action profiles of all agents excluding agent $k$, and $\otimes$ denote the Kronecker product. 
\begin{theorem}
\label{theorem:discrete-continuous}
Consider the interpolated process $\globalregretinterpol\cd$ defined in~\eqref{eq:global-regret-interpol}. Then, as $\stepsize\to 0$,
$\globalregretinterpol\cd$ converges weakly\footnote{Let $Z_\dtimee$ and $Z$ be $\RR^r$-valued random vectors. $Z_\dtimee$ converges weakly to $Z$, denoted by $Z_\dtimee \Rightarrow Z$, if for any bounded and continuous function $\psi(\cdot)$, $\ee\psi(Z_{\dtimee})\to \ee\psi(Z)$ as $\dtimee\to \infty.$} to $\globalregret\cd$ that is a solution of the system of interconnected differential inclusions
\begin{equation}
\label{eq:thrm-3-1}
{d\globalregret\over dt} \in \diffinclglobal\left(\globalregret\right) + (\mathbf{C}-I) \globalregret,
\end{equation}
where $\mathbf{C} = C \otimes I_{\fact}$ (see~\eqref{eq:W}), $\fact$ denotes the cardinality of the agents' action set, and
\begin{equation}
\label{eq:thrm-3-2}
\begin{split}
&h_{ij}\left(\globalregret\right) = \left\{\left[u^\kappa(\iota,\mixedstrat^{-\kappa}) - u^\kappa(j,\mixedstrat^{-\kappa})\right]\psi_{\iota}^\kappa; \mixedstrat^{-\kappa}\in\Delta\actset^{-\kappa}\right\},\\
&\hspace{2cm}\textstyle \iota = i\bmod \fact,\ \kappa = \left\lfloor {i\over \fact}\right\rfloor+1.\nonumber
\end{split}
\end{equation}
Further, $\statdistk$ represents the stationary distribution characterized in Lemma~\ref{lemma:stationary-distr}.
\begin{IEEEproof}
The proof relies on stochastic averaging theory and is omitted for brevity. The interested reader is referred to~\cite[Appendix~B]{KNH14} for a detailed proof.
\end{IEEEproof}
\end{theorem}


\subsubsection*{Step 2} Next, we examine stability of the limit system~\eqref{eq:thrm-3-1}, and show its set of global attractors comprises an $\epsilon$-neighborhood of the negative orthant. 
With slight abuse of notation, 
we rearrange the elements of the global regret matrix $\globalregret$ as a vector, but still denote it by $\globalregret$.
\begin{theorem}
\label{theorem:game-stability}
Consider the limit dynamical system~\eqref{eq:thrm-3-1}. Let
\begin{equation}
\label{eq:game-R}
\RR_-^{\CEdistance} = \lbr \x\in\RR^{\fplyr(\fact)^2}\; ;\; |\x|^+ \leq \CEdistance\mathbf{1}\rbr.
\end{equation}
Let further $\regmatplyr(0) = \regmatplyr_0$. Then, for each $\CEdistance \geq 0$, there exists $\widehat{\explor}\left(\CEdistance\right)\geq 0$ such that if $\explor\leq\widehat{\explor}\left(\CEdistance\right)$ in~\eqref{eq:statdistr-groups} (or equivalently in the decision strategy~\eqref{eq:strategy-MC-homog-social-group}), the set $\RR^\epsilon_-$ is globally asymptotically stable for the limit system~\eqref{eq:thrm-3-1}. That is,
\begin{equation}
\lim_{t\to\infty} \mathrm{dist}\lb \globalregret(t),\RR^\CEdistance_-\rb = 0,
\end{equation}
where $\mathrm{dist}\lb\cdot,\cdot\rb$ denotes the usual distance function.
\begin{IEEEproof}
The proof uses Lyapunov stability theory and is omitted for brevity. 
\end{IEEEproof}
\end{theorem}

Subsequently, we study asymptotic stability by looking at the case where $\stepsize\to 0$, $\dtimee\to\infty$, and $\stepsize \dtimee \to\infty$. Nevertheless, instead of considering a two-stage limit by first letting $\stepsize\rightarrow0$ and then $t\rightarrow\infty$, we study $\globalregret^{\stepsize}(t+t_{\stepsize})$ and require $t_{\stepsize}\rightarrow\infty$ as $\stepsize\rightarrow 0$. The following corollary asserts that the results of Theorem~\ref{theorem:game-stability} also hold for the interpolated processes. 
%
\begin{corollary}
\label{corollary:game-asymptotic-stab}
Denote by $\left\lbrace t_\stepsize\right\rbrace$ any sequence of real numbers satisfying $t_\mu\rightarrow\infty$ as $\mu\rightarrow0$. Suppose $\{\globalregret_\dtimee: \stepsize > 0, \dtimee <\infty\}$ is tight or bounded in probability. Then, for each $\epsilon \geq 0$, there exists $\widehat{\delta}\left(\CEdistance\right)\geq 0$ such that if $0<\explor\leq\widehat{\explor}(\CEdistance)$ in~\eqref{eq:strategy-MC-homog-social-group}, $\globalregret^{\stepsize}(\cdot+t_\stepsize)\to \RR^\CEdistance_- $ in probability, where $\RR^\CEdistance_-$ is defined in~\eqref{eq:game-R}.
\end{corollary}

The above corollary completes the proof of the first result in Theorem~\ref{theorem:main-game}.

\subsubsection*{Step 3} In the final step, we show that the convergence of the regrets of individual agents to an $\CEdistance$-neighborhood of the negative orthant provides the necessary and sufficient condition for convergence of their global behavior to the correlated $\CEdistance$-equilibria set. This is summarized in the following theorem.
\begin{theorem}
\label{theorem:game-CE-convergence}
Recall the interpolated processes for the global regret matrix $\globalregretinterpol\cd$, defined in~\eqref{eq:global-regret-interpol}, and agent's collective behavior $\globbehav^{\stepsize}\cd$, defined in~\eqref{eq:interpolated-game}. Then, $\globbehav^{\stepsize}\cd$ converges in probability to the correlated $\CEdistance$-equilibrium if and only if $\globalregretinterpol\cd\to \RR_-^\epsilon$ in probability, where $\RR_-^\epsilon$ is defined in~\eqref{eq:game-R}.
\begin{IEEEproof}
The proof relies on how the regrets are defined. The interested reader is referred to~\cite[Section~IV-D]{NKY13a} for a somewhat similar proof.
\end{IEEEproof}
\end{theorem}

The above theorem, together with Corollary~\ref{corollary:game-asymptotic-stab}, completes the proof for the second result in Theorem~\ref{theorem:main-game}.

\section{Proof of Theorem~\ref{thrm:Type-I_ErrorProbability}}
Consider a dataset $\dataset$ (\ref{eqn:ResponseData}) that satisfies Nash rationality (\ref{eqn:NashRationFesTest}). Given $\dataset$, the inequalities (\ref{eqn:NashRationFesTest}) have a feasible solution. Denote the solution parameters of (\ref{eqn:NashRationFesTest}), given $\dataset$, by $\{\lambda_t^{io}>0,V_t^o\}$. Substituting $\response_t^i=\obsresponse_t^i-\noise_t^i$, from (\ref{eqn:ObservationData}), into the inequalities obtained from the solution of (\ref{eqn:NashRationFesTest}) given $\dataset$, we obtain the inequalities:
	\begin{equation}
	V_{\tau}^o-V_t^o-\sum\limits_{i=1}^n\lambda_t^{io}\probe_\tindx^\p(\obsresponse_{\tau}^i-\obsresponse_t^i) \leq \sum\limits_{i=1}^n\lambda_t^{io}\probe_\tindx^\p(\noise_{t}^i-\noise_{\tau}^i).
	\label{eqn:cleanNLP}
	\end{equation}
	The goal is to compute an upper bound on the r.h.s. of (\ref{eqn:cleanNLP}) that is independent of $\lambda_t^{io}$. Notice that the following inequalities provide an upper bound on the r.h.s. of (\ref{eqn:cleanNLP}):
	\begin{align}
	\sum\limits_{i=1}^n\lambda_t^{io}\probe_\tindx^\p(\noise_{t}^i-\noise_{\tau}^i) &\leq \sum\limits_{i=1}^n\lambda_t^{io}|\probe_\tindx^\p(\noise_{t}^i-\noise_{\tau}^i)| \nonumber\\
	&\leq \Big(\sum\limits_{i=1}^n\lambda_t^{io}\Big)\Big(\sum\limits_{i=1}^n|\probe_\tindx^\p(\noise_{t}^i-\noise_{\tau}^i)|\Big) \nonumber\\
	&\leq \Lambda_\tindx M
	\label{eqn:inequaltybound}
	\end{align}
	with $\Lambda_\tindx=\sum_{i=1}^n\lambda_t^{io}$, and $M$ defined by (\ref{eqn:UpperBoundNLP}). Substituting (\ref{eqn:inequaltybound}) into (\ref{eqn:cleanNLP}) the following inequalities are obtained:
	\begin{equation}
	\frac{1}{\Lambda_\tindx}\Big(V_{\tau}^o-V_t^o-\sum\limits_{i=1}^n\lambda_t^{io}\probe_\tindx^\p(\obsresponse_{\tau}^i-\obsresponse_t^i)\Big) \leq M.
	\label{eqn:noiseBoundNLP}
	\end{equation}
	
	A solution of (\ref{eqn:NLP_NashRation}) given $\obsdataset$, defined by (\ref{eqn:ObservationData}), is denoted by $\{\Phi^*\{{\bf\obsresponse}\},\lambda_t^{i*},V_t^*\}$. By comparing the inequalities obtained from the solution of (\ref{eqn:NLP_NashRation}) given $\obsdataset$, and the inequalities (\ref{eqn:noiseBoundNLP}), notice that $\{\Phi^*\{{\bf\obsresponse}\}=M,\lambda_t^{i*}=\lambda_t^{io},V_t^*=V_t^o\}$ is a feasible, but not necessarily optimal solution of (\ref{eqn:NLP_NashRation}) given $\obsdataset$. Therefore, for $\dataset$ satisfying malicious cooperation, it must be the case that $\Phi^*\{{\bf\obsresponse}\}\leq M$. This asserts, under the null hypothesis $H_0$, that $\Phi^*\{{\bf\obsresponse}\}$ is upper bounded by $M$. For a given $\Phi^*\{{\bf\obsresponse}\}$, the integral in (\ref{eqn:Statistical_Test}) is the probability of $\Phi^*\{{\bf\obsresponse}\}\leq M$; therefore, the conditional probability of rejecting $H_0$ when true is less then $\gamma$. 
	\qed


%
%

\ifCLASSOPTIONcaptionsoff
  \newpage
\fi



\bibliographystyle{IEEEtran}
\bibliography{IEEEabrv,ref_V12}

%
%
%
%
%
%
%

\end{document}